\documentclass[10pt]{article}
\usepackage{graphicx}
\usepackage{amsmath}
\usepackage{amssymb}
\usepackage{caption2}
\begin{document}
\bibliographystyle{prsty}
\begin{center}
{\large {\bf \sc{  Analysis of  the $P_c(4380)$ and $P_c(4450)$ as pentaquark states in the diquark model with QCD sum rules }}} \\[2mm]
Zhi-Gang Wang \footnote{E-mail:zgwang@aliyun.com.  }     \\
 Department of Physics, North China Electric Power University, Baoding 071003, P. R. China

\end{center}

\begin{abstract}
In this article, we construct the diquark-diquark-antiquark type interpolating currents,  and  study the masses and pole residues  of the $J^P={\frac{3}{2}}^-$ and ${\frac{5}{2}}^+$ hidden-charm pentaquark states   in details with the QCD sum rules by calculating the contributions of the vacuum condensates up to dimension-10 in the operator product expansion. In calculations,  we use the  formula $\mu=\sqrt{M^2_{P_c}-(2{\mathbb{M}}_c)^2}$  to determine  the energy scales of the QCD spectral densities.  The present predictions favor assigning  the $P_c(4380)$ and $P_c(4450)$ to be the  ${\frac{3}{2}}^-$ and ${\frac{5}{2}}^+$ pentaquark states, respectively.
\end{abstract}



\section{Introduction}

In 1964, Gell-Mann suggested  that  multiquark states beyond the  minimal quark contents $q\bar{q}$ and $qqq$  maybe exist \cite{Gell-Mann-1964},
a quantitative model for the tetraquark states with the quark contents $qq\bar{q}\bar{q}$ was developed by Jaffe using the MIT bag model in 1976 \cite{Jaffe-1977}. Latter, the five-quark baryons with the quark contents $qqqq\bar{q}$ were developed \cite{Strottman-1979}, while the name pentaquark  was introduced by Lipkin \cite{Lipkin-1987}. The QCD  allows the existence of multiquark states and hybrid states which contain not
only quarks but also gluonic degrees of freedom.  We can construct the tetraquark states and pentaquark states according to the diquark-antidiquark model and diquark-diquark-antiquark model, respectively \cite{Maiani-2004,Jaffe-2003}.  In the light quark sector, the nature of the  scalar
 mesons  below  $1\,\rm{GeV}$ is under controversy  \cite{ReviewAmsler2}, although those light tetraquark states  are not ruled out in the $N_c$ limit \cite{Weinberg}.
 In the heavy quark sector, several $X$, $Y$ and $Z$ mesons are observed, such as the $Z_c(3900)^{\pm}$, $Z_c(4020/4025)^{\pm}$, $Z(4430)^{\pm}$, the net charge indicates that their constituents are $c\bar{c}u\bar{d}$ or $c\bar{c}d\bar{u}$, for recent review on both the experimental and theoretical aspects, one can consult Ref.\cite{Polosa-IJMPLA}. Some $X$, $Y$ and $Z$ mesons  are assigned tentatively to be tetraquark states, irrespective of the diquark-antidiquark type or
  the meson-meson type. The two  heavy quarks play an important role
  in stabilizing the multiquark systems, just as in the case
of the $(\mu^-e^+)(\mu^+ e^-)$ molecule in QED \cite{Brodsky-2014}.
 The spacial  separation between the diquark and antidiquark in the tetraquark states \cite{Brodsky-2014,Maiani-1405} (or meson and meson in the molecular states \cite{Voloshin-2011, Karliner-2015}) may lead to small decay widths, we can study the decay patterns  by performing   the Fierz rearrangements   non-relativistically   in the Pauli-spinor pace \cite{Maiani-1405,Voloshin-2011, Karliner-2015} or relativistically in the Dirac-spinor space \cite{Wang-Tetraquark-DW}.

Recently,  the  LHCb collaboration  observed  two exotic structures ($P_c(4380)$ and $P_c(4450)$) in the $J/\psi p$ mass spectrum in the $\Lambda_b^0\to J/\psi K^- p$ decays,  which are referred to be  charmonium-pentaquark states now \cite{LHCb-4380}.  The  $P_c(4380)$ has a mass of $4380\pm 8\pm 29\,\rm{MeV}$  and a width of $205\pm 18\pm 86\,\rm{MeV}$, while the $P_c(4450)$ has a mass of $4449.8\pm 1.7\pm 2.5\,\rm{MeV}$ and a width of $39\pm 5\pm 19\,\rm{MeV}$. The preferred spin-parity assignments of the $P_c(4380)$ and $P_c(4450)$ are  $J^P={\frac{3}{2}}^-$ and ${\frac{5}{2}}^+$, respectively.
The significance of each of the two resonances is more than $9\,\sigma$ \cite{LHCb-4380}.   The $P_c(4380)$ and $P_c(4450)$ have attracted  much attentions of the theoretical physicists, several  attempted assignments are suggested, such as the $\Sigma_c \bar{D}^*$, $\Sigma_c^* \bar{D}^*$, $\chi_{c1}p$ molecular pentaquark  states \cite{mole-penta} (or not the molecular pentaquark states \cite{mole-penta-No}),  the diquark-diquark-antiquark type pentaquark states \cite{di-di-anti-penta}, the diquark-triquark  type  pentaquark states \cite{di-tri-penta}, re-scattering effects \cite{rescattering-penta}, etc. We can test their resonant nature by using photoproduction off a proton target \cite{Test-Penta}.

The quarks have color $SU(3)$ symmetry,  we can construct the pentaquark states according to the routine  ${\rm quark}\to {\rm diquark}\to {\rm pentaquark}$,
\begin{eqnarray}
(3\otimes 3)\otimes(3\otimes 3) \otimes\overline{3}&=&(\overline{3}\oplus 6)\otimes(\overline{3}\oplus 6)\otimes\overline{3} =\overline{3}\otimes\overline{3}\otimes\overline{3}\oplus\cdots =1\oplus  \cdots \, ,
\end{eqnarray}
or construct the molecular pentaquark states according to the routine ${\rm quark}\to {\rm meson\,\, and\,\, baryon}$ $\to {\rm molecular\,\, pentaquark\,\,  state}$,
\begin{eqnarray}
(3\otimes \overline{3})\otimes(3\otimes 3\otimes 3)&=&(1\oplus 8)\otimes(1\oplus \cdots) =(1\otimes1)\oplus  \cdots=1\oplus  \cdots    \, ,
\end{eqnarray}
where the $1$, $3$ ($\overline{3}$), $6$  and $8$ denote the color singlet, triplet (antitriplet), sextet  and octet, respectively.
In the diquark model, the pentaquark states consist of two diquarks and an antiquark, which are colored constituents, it is easy to form compact bound states due to the
strong attractions at long distance.  In the meson-baryon model, the molecular pentaquark states consist of a colorless meson and a colorless baryon,    attractions induced by exchanges of the intermediate mesons (Yukawa-like potentials) are needed to form loose bound states.
In this article, we take the $P_c(4380)$ and $P_c(4450)$ as the diquark-diquark-antiquark type pentaquark states, construct the interpolating currents consist of five quarks according to Eq.(1),  and study their masses and pole residues with the QCD sum rules.

In previous works, we described   the hidden charm (or bottom) four-quark systems  $q\bar{q}^{\prime}Q\bar{Q}$
by a double-well potential \cite{Wang-Tetraquark-DW, Wang-molecule,Wang-Octet}.     In the four-quark system $q\bar{q}^{\prime}Q\bar{Q}$,
 the $Q$-quark serves as a static well potential and  combines with the light quark $q$  to form a heavy diquark $\mathcal{D}^i_{qQ}$ in  color antitriplet \cite{Wang-Tetraquark-DW},
\begin{eqnarray}
q+Q &\to & \mathcal{D}^i_{qQ} \, ,
\end{eqnarray}
or combines with the light antiquark $\bar{q}^\prime$ to form a heavy meson in color singlet (meson-like state in color octet) \cite{Wang-molecule,Wang-Octet}
\begin{eqnarray}
\bar{q}^\prime+Q &\to & \bar{q}^{\prime}Q\,\, (\bar{q}^{\prime}\lambda^{a}Q) \, ,
\end{eqnarray}
 the $\bar{Q}$-quark serves  as another static well potential and combines with the light antiquark $\bar{q}^\prime$  to form a heavy antidiquark $\mathcal{D}^i_{\bar{q}^{\prime}\bar{Q}}$ in  color triplet \cite{Wang-Tetraquark-DW},
\begin{eqnarray}
\bar{q}^{\prime}+\bar{Q} &\to & \mathcal{D}^i_{\bar{q}^{\prime}\bar{Q}} \, ,
\end{eqnarray}
or combines with the light quark $q$ to form a heavy meson in color singlet (meson-like state in color octet) \cite{Wang-molecule,Wang-Octet}
\begin{eqnarray}
q+\bar{Q} &\to & \bar{Q}q\,\, (\bar{Q}\lambda^{a}q) \, ,
\end{eqnarray}
where the $i$ is color index, the $\lambda^a$ is Gell-Mann matrix.
 Then
\begin{eqnarray}
 \mathcal{D}^i_{qQ}+\mathcal{D}^i_{\bar{q}^{\prime}\bar{Q}} &\to &  {\rm compact \,\,\, tetraquark \,\,\, states}\, , \nonumber\\
 \bar{q}^{\prime}Q+\bar{Q}q &\to & {\rm loose  \,\,\, molecular \,\,\, states}\, , \nonumber\\
  \bar{q}^{\prime}\lambda^aQ+\bar{Q}\lambda^a q &\to & {\rm   molecule-like  \,\,\, states}\, ,
\end{eqnarray}
the two heavy quarks $Q$ and $\bar{Q}$ stabilize the four-quark systems $q\bar{q}^{\prime}Q\bar{Q}$, just as in the case
of the $(\mu^-e^+)(\mu^+ e^-)$ molecule in QED \cite{Brodsky-2014}.

The hidden charm (or bottom) five-quark systems  $qq_1q_2Q\bar{Q}$ can  also  be described
by a double-well potential by using the replacement,
\begin{eqnarray}
q_1+q_2 +\bar{Q}&\to & \mathcal{D}_{q_1q_2\,(\bar{q}^{\prime  })}^j +\bar{Q}^k \to  \mathcal{T}^i_{q_1q_2\,(\bar{q}^\prime)\bar{Q}} \, ,
\end{eqnarray}
 just like the four-quark systems $q\bar{q}^{\prime}Q\bar{Q}$ \cite{Wang-Tetraquark-DW, Wang-molecule}, where the $\mathcal{T}^i_{q_1q_2\bar{Q}}$ denotes the heavy
 triquark in color triplet,  the  $\bar{q}^\prime$ in the bracket denotes that the $\mathcal{D}_{q_1q_2}^j$ is in  color antitriplet, just like the $\bar{q}^{\prime j}$.
In the heavy quark limit, the $Q$-quark ($\bar{Q}$-quark) can be taken as a static well potential, the diquark $\mathcal{D}_{q_1q_2}^j$ and quark $q$  lie in the two wells,  respectively.

The QCD sum rules have been applied extensively  to study the  hidden-charm (bottom) tetraquark states \cite{No-formular}, however, the energy scale dependence  of the QCD spectral densities is not studied. In previous works, we studied the acceptable energy scales of the QCD spectral densities  for the hidden  charm (bottom) tetraquark states and molecular (and molecule-like) states in the QCD sum rules in details for the first time \cite{Wang-Tetraquark-DW, Wang-molecule,Wang-Octet,WangHuang-PRD,Wang-Tetraquark-DW-2},  and suggested a  formula
 \begin{eqnarray}
 \mu&=&\sqrt{M^2_{X/Y/Z}-(2{\mathbb{M}}_Q)^2}\, ,
  \end{eqnarray}
  to determine  the energy scales based on the analysis in Eqs.(3-7), where the $X$, $Y$, $Z$ denote the four-quark systems, and the ${\mathbb{M}}_Q$ denotes the effective heavy quark masses  \cite{Wang-Tetraquark-DW, Wang-molecule,Wang-Octet}. The energy scale formula works well for all   the tetraquark states, molecular states and molecule-like states.

  In the non-relativistic quark model, the heavy quarks have finite masses, which quantitatively  affect  the spin-spin interactions between the quarks
 within one diquark or in two different diquarks \cite{Maiani-1405}.
 In the QCD sum rules,   the net effects of the  different dynamics are embodied in the effect masses ${\mathbb{M}}_c$ and ${\mathbb{M}}_b$, respectively, for example,
 the $Z_c(3900)$ and $Z_b(10610)$ can be tentatively assigned to be the $J^{PC}=1^{+-}$ tetraquark states with the symbolic quark structures $\frac{[cu]_{S=0}[\bar{c}\bar{d}]_{S=1} - [cu]_{S=1}[\bar{c}\bar{d}]_{S=0}}{\sqrt{2}}$ and $\frac{[bu]_{S=0}[\bar{b}\bar{d}]_{S=1} - [bu]_{S=1}[\bar{b}\bar{d}]_{S=0}}{\sqrt{2}}$, respectively, where the subscript $S$ denotes the spin,  the optimal energy scales of their QCD spectral densities are quite different $\mu_{Z_c(3900)}=1.5\,\rm{GeV}$ and $\mu_{Z_b(10610)}=2.7\,\rm{GeV}$ \cite{Wang-Tetraquark-DW, WangHuang-PRD}, although they are cousins. While in the heavy quark limit $m_Q \to  \infty$, we naively expect that the two energy scales $\mu_{Z_c(3900)}$ and $\mu_{Z_b(10610)}$ coincide.
In this work, we extend the energy scale formula to study the diquark-diquark-antiquark type pentaquark states, and try to assign the $P_c(4380)$ and $P_c(4450)$ to be the ${\frac{3}{2}}^-$ and ${\frac{5}{2}}^+$ pentaquark states, respectively.

 The article is arranged as follows:  we derive the QCD sum rules for the masses and pole residues of  the
$P_c(4380)$ and $P_c(4450)$ in Sect.2;  in Sect.3, we present the numerical results and discussions; and Sect.4 is reserved for our
conclusions.

\section{QCD sum rules for  the $P_c(4380)$ and $P_c(4450)$}

In the following, we write down  the two-point correlation functions $\Pi_{\mu\nu}(p)$ and $\Pi_{\mu\nu\alpha\beta}(p)$  in the QCD sum rules,
\begin{eqnarray}
\Pi_{\mu\nu}(p)&=&i\int d^4x e^{ip \cdot x} \langle0|T\left\{J_{\mu}(x)\bar{J}_{\nu}(0)\right\}|0\rangle \, , \\
\Pi_{\mu\nu\alpha\beta}(p)&=&i\int d^4x e^{ip \cdot x} \langle0|T\left\{J_{\mu\nu}(x)\bar{J}_{\alpha\beta}(0)\right\}|0\rangle \, ,
\end{eqnarray}
where
\begin{eqnarray}
 J_{\mu}(x)&=&\varepsilon^{ila} \varepsilon^{ijk}\varepsilon^{lmn}  u^T_j(x) C\gamma_5 d_k(x)\,u^T_m(x) C\gamma_\mu c_n(x)\, C\bar{c}^{T}_{a}(x) \, , \\
 J_{\mu\nu}(x)&=&\frac{1}{\sqrt{2}}\varepsilon^{ila} \varepsilon^{ijk}\varepsilon^{lmn}  u^T_j(x) C\gamma_5 d_k(x)\left[u^T_m(x) C\gamma_\mu c_n(x)\, \gamma_{\nu}C\bar{c}^{T}_{a}(x)+u^T_m(x) C\gamma_\nu c_n(x)\, \gamma_{\mu}C\bar{c}^{T}_{a}(x)\right] \, ,\nonumber\\
\end{eqnarray}
the $i$, $j$, $k$, $\cdots$ are color indices, the $C$ is the charge conjugation matrix.
The diquarks $q^{T}_j C\Gamma q^{\prime}_k$ have  five  structures  in Dirac spinor space, where $C\Gamma=C\gamma_5$, $C$, $C\gamma_\mu \gamma_5$,  $C\gamma_\mu $ and $C\sigma_{\mu\nu}$ for the scalar, pseudoscalar, vector, axialvector  and  tensor diquarks, respectively.  The structures
$C\gamma_\mu $ and $C\sigma_{\mu\nu}$ are symmetric, while the structures
$C\gamma_5$, $C$ and $C\gamma_\mu \gamma_5$ are antisymmetric.
The scattering amplitude for one-gluon exchange  is proportional to
\begin{eqnarray}
\left(\frac{\lambda^a}{2}\right)_{ki}\left(\frac{\lambda^a}{2}\right)_{lj}&=&-\frac{1}{3}\left(\delta_{jk}\delta_{il}-\delta_{ik}\delta_{jl}\right)
 +\frac{1}{6}\left(\delta_{jk}\delta_{il}+\delta_{ik}\delta_{jl}\right)\, ,
\end{eqnarray}
where  the $i,j$ and $k,l$ are the color indexes of the two quarks in the incoming
and outgoing channels respectively.   The negative sign in front of the antisymmetric  antitriplet indicates the interaction
is attractive  while the positive sign in front of the symmetric
sextet indicates  the interaction  is repulsive.
The
attractive interactions of one-gluon exchange  favor  formation of
the diquarks in  color antitriplet $\overline{3}_{ c}$, flavor
antitriplet $\overline{3}_{ f}$ and spin singlet $1_s$ \cite{One-gluon},
 while the favored configurations are the scalar ($C\gamma_5$) and axialvector ($C\gamma_\mu$) diquark states \cite{WangDiquark,WangLDiquark}.
  The calculations based on the QCD sum rules indicate that the  heavy-light scalar and axialvector  diquark states have almost  degenerate masses \cite{WangDiquark},  while  the masses of the
 light  axialvector  diquark states lie   $(150-200)\,\rm{MeV}$ above that of the  light scalar diquark states \cite{WangLDiquark}, if they have the same quark constituents. In this article, we choose the light scalar diquark and heavy axialvector diquark as basic constituents,
    and construct  the scalar-diquark-axialvector-diquark-antiquark type currents $J_{\mu}(x)$ and $J_{\mu\nu}$ with the spin-parity ${\frac{3}{2}}^-$ and ${\frac{5}{2}}^+$ respectively to interpolate    the  pentaquark states $P_c(4380)$ and $P_c(4450)$, respectively, see Eq.(3) and Eq.(8).

 In fact, we can also construct the axialvector-diquark-scalar-diquark-antiquark type current $\eta_{\mu}(x)$ and axialvector-diquark-axialvector-diquark-antiquark type current $\eta_{\mu\nu}(x)$,
\begin{eqnarray}
 \eta_{\mu}(x)&=&\frac{\varepsilon^{ila} \varepsilon^{ijk}\varepsilon^{lmn}}{\sqrt{3}} \left[ u^T_j(x) C\gamma_\mu u_k(x) d^T_m(x) C\gamma_5 c_n(x)+2u^T_j(x) C\gamma_\mu d_k(x) u^T_m(x) C\gamma_5 c_n(x)\right]    C\bar{c}^{T}_{a}(x) \, ,  \nonumber\\
 \eta_{\mu\nu}(x)&=&\frac{\varepsilon^{ila} \varepsilon^{ijk}\varepsilon^{lmn}}{\sqrt{6}} \left[ u^T_j(x) C\gamma_\mu u_k(x)d^T_m(x) C\gamma_\nu c_n(x)+2u^T_j(x) C\gamma_\mu d_k(x)u^T_m(x) C\gamma_\nu c_n(x) \right]\gamma_5 C\bar{c}^{T}_{a}(x)   \nonumber\\
 &&+\left( \mu\leftrightarrow\nu\right)\, ,
 \end{eqnarray}
 to study the spin-parity ${\frac{3}{2}}^-$ and ${\frac{5}{2}}^+$ pentaquark states, respectively.
As the masses of the  light  axialvector  diquark states lie   $(150-200)\,\rm{MeV}$ above that of the corresponding  light scalar diquark states \cite{WangLDiquark}.
The currents $\eta_\mu(x)$ and $\eta_{\mu\nu}(x)$ are supposed to couple to the pentaquark states with larger masses compared to the currents $J_\mu(x)$ and $J_{\mu\nu}(x)$, respectively.

The $\Lambda_b^0$ can be well interpolated by the current $J(x)=\varepsilon^{ijk}u^T_i(x) C\gamma_5 d_j(x) b_k(x)$ \cite{WangLambda}, the $u$ and $d$ quarks in the $\Lambda_b^0$ form a scalar diquark $[ud]$ in color antitriplet, the decays $\Lambda_b^0\to J/\psi  p K^-$ take  place through the mechanism,
 \begin{eqnarray}
 \Lambda_b^0([ud] b )&\to& [ud] c\bar{c}s\to [ud] c\bar{c}u\bar{u}s\to P_c^+([ud] [uc]\bar{c})K^-(\bar{u}s)\to J/\psi  p K^- \, ,
  \end{eqnarray}
  at the quark level.  In the decays $P_c^+([ud] [uc]\bar{c})\to J/\psi  p$, the scalar diquark $[ud]$ survives in the decays, the decays are greatly facilitated. On the other hand, if there exists a light axialvector diquark $[ud]$, which has to dissolve to form a scalar diquark $[ud]$, the    decays are not facilitated.

The currents $J_\mu(0)$ and $J_{\mu\nu}(0)$ couple potentially to the ${\frac{1}{2}}^+$, ${\frac{3}{2}}^-$ and ${\frac{1}{2}}^+$, ${\frac{3}{2}}^-$, ${\frac{5}{2}}^+$  hidden-charm  pentaquark
 states $P_{\frac{1}{2}}^{+}$, $P_{\frac{3}{2}}^{-}$ and $P_{\frac{1}{2}}^{+}$, $P_{\frac{3}{2}}^{-}$, $P_{\frac{5}{2}}^{+}$, respectively,
\begin{eqnarray}
\langle 0| J_{\mu} (0)|P_{\frac{1}{2}}^{+}(p)\rangle &=&f^{+}_{\frac{1}{2}}p_\mu U^{+}(p,s) \, , \nonumber \\
\langle 0| J_{\mu} (0)|P_{\frac{3}{2}}^{-}(p)\rangle &=&\lambda^{-}_{\frac{3}{2}} U^{-}_\mu(p,s) \, ,  \\
\langle 0| J_{\mu\nu} (0)|P_{\frac{1}{2}}^{+}(p)\rangle &=&g^{+}_{\frac{1}{2}}p_\mu p_\nu U^{+}(p,s) \, , \nonumber\\
\langle 0| J_{\mu\nu} (0)|P_{\frac{3}{2}}^{-}(p)\rangle &=&f^{-}_{\frac{3}{2}} \left[p_\mu U^{-}_{\nu}(p,s)+p_\nu U^{-}_{\mu}(p,s)\right] \, , \nonumber\\
\langle 0| J_{\mu\nu} (0)|P_{\frac{5}{2}}^{+}(p)\rangle &=&\lambda^{+}_{\frac{5}{2}} U^{+}_{\mu\nu}(p,s) \, ,
\end{eqnarray}
the spinors $U^\pm(p,s)$ satisfy the Dirac equations  $(\not\!\!p-M_{\pm})U^{\pm}(p)=0$, while the spinors $U^{\pm}_\mu(p,s)$ and $U^{\pm}_{\mu\nu}(p,s)$ satisfy the Rarita-Schwinger equations $(\not\!\!p-M_{\pm})U^{\pm}_\mu(p)=0$ and $(\not\!\!p-M_{\pm})U^{\pm}_{\mu\nu}(p)=0$,  and the relations $\gamma^\mu U^{\pm}_\mu(p,s)=0$,
$p^\mu U^{\pm}_\mu(p,s)=0$, $\gamma^\mu U^{\pm}_{\mu\nu}(p,s)=0$,
$p^\mu U^{\pm}_{\mu\nu}(p,s)=0$, $ U^{\pm}_{\mu\nu}(p,s)= U^{\pm}_{\nu\mu}(p,s)$, respectively.   On the other hand, the currents $J_\mu(0)$ and $J_{\mu\nu}(0)$ also couple potentially to the ${\frac{1}{2}}^-$, ${\frac{3}{2}}^+$ and ${\frac{1}{2}}^-$, ${\frac{3}{2}}^+$, ${\frac{5}{2}}^-$  hidden-charm  pentaquark
 states $P_{\frac{1}{2}}^{-}$, $P_{\frac{3}{2}}^{+}$ and $P_{\frac{1}{2}}^{-}$, $P_{\frac{3}{2}}^{+}$, $P_{\frac{5}{2}}^{-}$, respectively,
 \begin{eqnarray}
\langle 0| J_{\mu} (0)|P_{\frac{1}{2}}^{-}(p)\rangle &=&f^{-}_{\frac{1}{2}}p_\mu i\gamma_5 U^{-}(p,s) \, , \nonumber\\
\langle 0| J_{\mu} (0)|P_{\frac{3}{2}}^{+}(p)\rangle &=&\lambda^{+}_{\frac{3}{2}}i\gamma_5 U^{+}_{\mu}(p,s) \, , \\
\langle 0| J_{\mu\nu} (0)|P_{\frac{1}{2}}^{-}(p)\rangle &=&g^{-}_{\frac{1}{2}}p_\mu p_\nu i\gamma_5 U^{-}(p,s) \, , \nonumber\\
\langle 0| J_{\mu\nu} (0)|P_{\frac{3}{2}}^{+}(p)\rangle &=&f^{+}_{\frac{3}{2}} i\gamma_5\left[p_\mu U^{+}_{\nu}(p,s)+p_\nu U^{+}_{\mu}(p,s)\right] \, , \nonumber\\
\langle 0| J_{\mu\nu} (0)|P_{\frac{5}{2}}^{-}(p)\rangle &=&\lambda^{-}_{\frac{5}{2}}i\gamma_5 U^{-}_{\mu\nu}(p,s) \, ,
\end{eqnarray}
the spinors $U^{-}_\mu(p,s)$ and $U^{+}_\mu(p,s)$ ($U^{-}_{\mu\nu}(p,s)$ and $U^{+}_{\mu\nu}(p,s)$)  have analogous  properties,  and the pole residues $\lambda^{\pm}_{\frac{3}{2}/\frac{5}{2}}\neq 0$, $f^{\pm}_{\frac{1}{2}/\frac{3}{2}}\neq 0$ and $g^{\pm}_{\frac{1}{2}}\neq 0$.

 We  insert  a complete set  of intermediate pentaquark states with the
same quantum numbers as the current operators $J_\mu(x)$,
$i\gamma_5 J_\mu(x)$, $J_{\mu\nu}(x)$ and
$i\gamma_5 J_{\mu\nu}(x)$ into the correlation functions
$\Pi_{\mu\nu}(p)$ and $\Pi_{\mu\nu\alpha\beta}(p)$ to obtain the hadronic representation
\cite{SVZ79,PRT85}. After isolating the pole terms of the lowest
states of the hidden-charm  pentaquark states, we obtain the
following results:
\begin{eqnarray}
   \Pi_{\mu\nu}(p) & = & {\lambda^{-}_{\frac{3}{2}}}^2  {\!\not\!{p}+ M_{-} \over M_{-}^{2}-p^{2}  } \left(- g_{\mu\nu}+\frac{\gamma_\mu\gamma_\nu}{3}+\frac{2p_\mu p_\nu}{3p^2}-\frac{p_\mu\gamma_\nu-p_\nu \gamma_\mu}{3\sqrt{p^2}}
\right)\nonumber\\
&&+  {\lambda^{+}_{\frac{3}{2}}}^2  {\!\not\!{p}- M_{+} \over M_{+}^{2}-p^{2}  } \left(- g_{\mu\nu}+\frac{\gamma_\mu\gamma_\nu}{3}+\frac{2p_\mu p_\nu}{3p^2}-\frac{p_\mu\gamma_\nu-p_\nu \gamma_\mu}{3\sqrt{p^2}}
\right)   \nonumber \\
& &+ {f^{+}_{\frac{1}{2}}}^2  {\!\not\!{p}+ M_{+} \over M_{+}^{2}-p^{2}  } p_\mu p_\nu+  {f^{-}_{\frac{1}{2}}}^2  {\!\not\!{p}- M_{-} \over M_{-}^{2}-p^{2}  } p_\mu p_\nu  +\cdots  \, ,\\
\Pi_{\mu\nu\alpha\beta}(p) & = & {\lambda^{+}_{\frac{5}{2}}}^2  {\!\not\!{p}+ M_{+} \over M_{+}^{2}-p^{2}  } \left[\frac{ \widetilde{g}_{\mu\alpha}\widetilde{g}_{\nu\beta}+\widetilde{g}_{\mu\beta}\widetilde{g}_{\nu\alpha}}{2}-\frac{\widetilde{g}_{\mu\nu}\widetilde{g}_{\alpha\beta}}{5}-\frac{1}{10}\left( \gamma_{\mu}\gamma_{\alpha}+\frac{\gamma_{\mu}p_{\alpha}-\gamma_{\alpha}p_{\mu}}{\sqrt{p^2}}-\frac{p_{\mu}p_{\alpha}}{p^2}\right)\widetilde{g}_{\nu\beta}\right.\nonumber\\
&&\left.-\frac{1}{10}\left( \gamma_{\nu}\gamma_{\alpha}+\frac{\gamma_{\nu}p_{\alpha}-\gamma_{\alpha}p_{\nu}}{\sqrt{p^2}}-\frac{p_{\nu}p_{\alpha}}{p^2}\right)\widetilde{g}_{\mu\beta}
+\cdots\right]\nonumber\\
&&+   {\lambda^{-}_{\frac{5}{2}}}^2  {\!\not\!{p}- M_{-} \over M_{-}^{2}-p^{2}  } \left[\frac{ \widetilde{g}_{\mu\alpha}\widetilde{g}_{\nu\beta}+\widetilde{g}_{\mu\beta}\widetilde{g}_{\nu\alpha}}{2}
-\frac{\widetilde{g}_{\mu\nu}\widetilde{g}_{\alpha\beta}}{5}-\frac{1}{10}\left( \gamma_{\mu}\gamma_{\alpha}+\frac{\gamma_{\mu}p_{\alpha}-\gamma_{\alpha}p_{\mu}}{\sqrt{p^2}}-\frac{p_{\mu}p_{\alpha}}{p^2}\right)\widetilde{g}_{\nu\beta}\right.\nonumber\\
&&\left.
-\frac{1}{10}\left( \gamma_{\nu}\gamma_{\alpha}+\frac{\gamma_{\nu}p_{\alpha}-\gamma_{\alpha}p_{\nu}}{\sqrt{p^2}}-\frac{p_{\nu}p_{\alpha}}{p^2}\right)\widetilde{g}_{\mu\beta}
 +\cdots\right]   \nonumber\\
 && +{f^{-}_{\frac{3}{2}}}^2  {\!\not\!{p}+ M_{-} \over M_{-}^{2}-p^{2}  } \left[ p_\mu p_\alpha \left(- g_{\nu\beta}+\frac{\gamma_\nu\gamma_\beta}{3}+\frac{2p_\nu p_\beta}{3p^2}-\frac{p_\nu\gamma_\beta-p_\beta \gamma_\nu}{3\sqrt{p^2}}
\right)+\cdots \right]\nonumber\\
&&+  {f^{+}_{\frac{3}{2}}}^2  {\!\not\!{p}- M_{+} \over M_{+}^{2}-p^{2}  } \left[ p_\mu p_\alpha \left(- g_{\nu\beta}+\frac{\gamma_\nu\gamma_\beta}{3}+\frac{2p_\nu p_\beta}{3p^2}-\frac{p_\nu\gamma_\beta-p_\beta \gamma_\nu}{3\sqrt{p^2}}
\right)+\cdots \right]   \nonumber \\
& &+ {g^{+}_{\frac{1}{2}}}^2  {\!\not\!{p}+ M_{+} \over M_{+}^{2}-p^{2}  } p_\mu p_\nu p_\alpha p_\beta+  {g^{-}_{\frac{1}{2}}}^2  {\!\not\!{p}- M_{-} \over M_{-}^{2}-p^{2}  } p_\mu p_\nu p_\alpha p_\beta  +\cdots \, ,
    \end{eqnarray}
where $\widetilde{g}_{\mu\nu}=g_{\mu\nu}-\frac{p_{\mu}p_{\nu}}{p^2}$, the $M_{\pm}$ are the masses of the lowest pentaquark states with the
 parity $\pm$ respectively, and the $\lambda^{\pm}_{\frac{3}{2}/\frac{5}{2}}$, $f^{\pm}_{\frac{1}{2}/\frac{3}{2}}$ and $g^{\pm}_{\frac{1}{2}}$ are the
corresponding pole residues.
In calculations, we have used the following summations \cite{HuangShiZhong},
\begin{eqnarray}
\sum_s U_\mu \overline{U}_\nu&=&\left(\!\not\!{p}+M_{\pm}\right)\left( -g_{\mu\nu}+\frac{\gamma_\mu\gamma_\nu}{3}+\frac{2p_\mu p_\nu}{3p^2}-\frac{p_\mu
\gamma_\nu-p_\nu \gamma_\mu}{3\sqrt{p^2}} \right) \,  ,  \\
\sum_s U_{\mu\nu}\overline {U}_{\alpha\beta}&=&\left(\!\not\!{p}+M_{\pm}\right)\left\{\frac{\widetilde{g}_{\mu\alpha}\widetilde{g}_{\nu\beta}+\widetilde{g}_{\mu\beta}\widetilde{g}_{\nu\alpha}}{2} -\frac{\widetilde{g}_{\mu\nu}\widetilde{g}_{\alpha\beta}}{5}-\frac{1}{10}\left( \gamma_{\mu}\gamma_{\alpha}+\frac{\gamma_{\mu}p_{\alpha}-\gamma_{\alpha}p_{\mu}}{\sqrt{p^2}}-\frac{p_{\mu}p_{\alpha}}{p^2}\right)\widetilde{g}_{\nu\beta}\right. \nonumber\\
&&-\frac{1}{10}\left( \gamma_{\nu}\gamma_{\alpha}+\frac{\gamma_{\nu}p_{\alpha}-\gamma_{\alpha}p_{\nu}}{\sqrt{p^2}}-\frac{p_{\nu}p_{\alpha}}{p^2}\right)\widetilde{g}_{\mu\beta}
-\frac{1}{10}\left( \gamma_{\mu}\gamma_{\beta}+\frac{\gamma_{\mu}p_{\beta}-\gamma_{\beta}p_{\mu}}{\sqrt{p^2}}-\frac{p_{\mu}p_{\beta}}{p^2}\right)\widetilde{g}_{\nu\alpha}\nonumber\\
&&\left.-\frac{1}{10}\left( \gamma_{\nu}\gamma_{\beta}+\frac{\gamma_{\nu}p_{\beta}-\gamma_{\beta}p_{\nu}}{\sqrt{p^2}}-\frac{p_{\nu}p_{\beta}}{p^2}\right)\widetilde{g}_{\mu\alpha} \right\} \, ,
\end{eqnarray}
and $p^2=M^2_{\pm}$ on the mass-shell.

We can rewrite the correlation functions $\Pi_{\mu\nu}(p)$ and $\Pi_{\mu\nu\alpha\beta}(p)$ into the following form according to Lorentz covariance,
\begin{eqnarray}
\Pi_{\mu\nu}(p)&=&\Pi_{\frac{3}{2}}(p^2)\,\left(- g_{\mu\nu}\right)+\Pi_{\frac{3}{2}}^1(p^2)\,\gamma_\mu \gamma_\nu+\Pi_{\frac{3}{2}}^2(p^2)\,\left(p_\mu\gamma_\nu-p_\nu \gamma_\mu\right) +\Pi_{\frac{1}{2},\frac{3}{2}}(p^2)\, p_\mu p_\nu\, , \\
\Pi_{\mu\nu\alpha\beta}(p)&=&\Pi_{\frac{5}{2}}(p^2)\,\frac{ g_{\mu\alpha}g_{\nu\beta}+g_{\mu\beta}g_{\nu\alpha}}{2}+\Pi_{\frac{5}{2}}^1(p^2)\, g_{\mu\nu}g_{\alpha\beta}+\Pi_{\frac{5}{2}}^2(p^2)\, \left(g_{\mu\nu}p_{\alpha}p_{\beta}+g_{\alpha\beta}p_{\mu}p_{\nu}\right) \nonumber\\
&&+\Pi_{\frac{5}{2}}^3(p^2)\,\left(  g_{\mu \alpha} \gamma_\nu \gamma_\beta+ g_{\mu \beta} \gamma_\nu \gamma_\alpha+ g_{\nu \alpha} \gamma_\mu \gamma_\beta+ g_{\nu \beta} \gamma_\mu \gamma_\alpha \right) \nonumber\\
&&+\Pi_{\frac{5}{2}}^4(p^2)\,\left[  g_{\nu \beta}\left(\gamma_{\mu} p_{\alpha}- \gamma_{\alpha}p_{\mu}\right) +
g_{\nu \alpha}\left(\gamma_{\mu}p_{ \beta}-\gamma_{ \beta}p_{\mu}\right) + g_{\mu \beta}\left(\gamma_{\nu} p_{\alpha}- \gamma_{\alpha}p_{\nu}\right)\right.\nonumber\\
&&\left.+ g_{\mu \alpha} \left(\gamma_{\nu}p_{ \beta}-\gamma_{ \beta}p_{\nu} \right)\right]\nonumber\\
&&+\Pi_{\frac{3}{2},\frac{5}{2}}^1(p^2)\,\left(  g_{\mu \alpha} p_\nu p_\beta+ g_{\mu \beta} p_\nu p_\alpha+ g_{\nu \alpha} p_\mu p_\beta+ g_{\nu \beta} p_\mu p_\alpha \right) \nonumber\\
&&+\Pi_{\frac{3}{2},\frac{5}{2}}^2(p^2)\,\left(  \gamma_{\mu} \gamma_{\alpha} p_\nu p_\beta+ \gamma_{\mu}\gamma_{ \beta} p_\nu p_\alpha+ \gamma_{\nu} \gamma_{\alpha} p_\mu p_\beta+ \gamma_{\nu}\gamma_{ \beta} p_\mu p_\alpha \right) \nonumber\\
&&+\Pi_{\frac{3}{2},\frac{5}{2}}^3(p^2)\,\left[  \left(\gamma_{\mu} p_{\alpha}- \gamma_{\alpha}p_{\mu}\right) p_\nu p_\beta+
\left(\gamma_{\mu}p_{ \beta}-\gamma_{ \beta}p_{\mu}\right) p_\nu p_\alpha+ \left(\gamma_{\nu} p_{\alpha}- \gamma_{\alpha}p_{\nu}\right) p_\mu p_\beta\right.\nonumber\\
&&\left.+ \left(\gamma_{\nu}p_{ \beta}-\gamma_{ \beta}p_{\nu} \right)p_\mu p_\alpha \right] +\Pi_{\frac{1}{2},\frac{3}{2},\frac{5}{2}}(p^2)\,p_\mu p_\nu p_\alpha p_\beta \, ,
\end{eqnarray}
 the subscripts $\frac{1}{2}$, $\frac{3}{2}$ and $\frac{5}{2}$ in the components $\Pi_{\frac{3}{2}}(p^2)$, $\Pi_{\frac{3}{2}}^1(p^2)$, $\Pi_{\frac{3}{2}}^2(p^2)$,  $\Pi_{\frac{1}{2},\frac{3}{2}}(p^2)$, $\Pi_{\frac{5}{2}}(p^2)$, $\Pi_{\frac{5}{2}}^1(p^2)$, $\Pi_{\frac{5}{2}}^2(p^2)$, $\Pi_{\frac{5}{2}}^3(p^2)$, $\Pi_{\frac{5}{2}}^4(p^2)$, $\Pi_{\frac{3}{2},\frac{5}{2}}^1(p^2)$, $\Pi_{\frac{3}{2},\frac{5}{2}}^2(p^2)$, $\Pi_{\frac{3}{2},\frac{5}{2}}^3(p^2)$ and
$\Pi_{\frac{1}{2},\frac{3}{2},\frac{5}{2}}(p^2)$ denote the spins the pentaquark states, which means that  the pentaquark states with $J=\frac{1}{2}$, $\frac{3}{2}$ and $\frac{5}{2}$ have contributions. The components $\Pi_{\frac{1}{2},\frac{3}{2}}(p^2)$, $\Pi_{\frac{3}{2},\frac{5}{2}}^1(p^2)$, $\Pi_{\frac{3}{2},\frac{5}{2}}^2(p^2)$, $\Pi_{\frac{3}{2},\frac{5}{2}}^3(p^2)$ and $\Pi_{\frac{1}{2},\frac{3}{2},\frac{5}{2}}(p^2)$ receive contributions from more than one pentaquark state, so they can be neglected. We can rewrite $\gamma_\mu \gamma_\nu=g_{\mu\nu}-i\sigma_{\mu\nu}$, then the components $\Pi_{\frac{3}{2}}^1(p^2)$, $\Pi_{\frac{3}{2}}^2(p^2)$, $\Pi_{\frac{5}{2}}^3(p^2)$ and $\Pi_{\frac{5}{2}}^4(p^2)$ are associated with tensor structures which are antisymmetric in the Lorentz indexes  $\mu$, $\nu$, $\alpha$ or $\beta$. In calculations, we observe that such antisymmetric properties lead to smaller intervals of dimensions of the vacuum condensates, therefore worse QCD sum rules, so the components $\Pi_{\frac{3}{2}}^1(p^2)$, $\Pi_{\frac{3}{2}}^2(p^2)$, $\Pi_{\frac{5}{2}}^3(p^2)$ and $\Pi_{\frac{5}{2}}^4(p^2)$ can also be  neglected. If we take the replacement $J_{\mu\nu}(x)\to \widehat{J}_{\mu\nu}(x)=J_{\mu\nu}(x)-\frac{1}{4}g_{\mu\nu}J_\alpha{}^\alpha(x)$ to subtract the contributions of the $J=\frac{1}{2}$ pentaquark states, a lots of terms   $\propto g_{\mu\nu}$, $g_{\alpha\beta}$  disappear at the QCD side, and result in smaller intervals of dimensions of the vacuum condensates, so the components  $\Pi_{\frac{5}{2}}^1(p^2)$ and $\Pi_{\frac{5}{2}}^2(p^2)$ are not the optimal choices  to study the $J=\frac{5}{2}$ pentaquark states. Now only the components  $\Pi_{\frac{3}{2}}(p^2)$ and $\Pi_{\frac{5}{2}}(p^2)$ are left. The present conclusion is tentative, we can obtain definite conclusion by obtaining  QCD sum rules based on the components   $\Pi_{\frac{3}{2}}^1(p^2)$, $\Pi_{\frac{3}{2}}^2(p^2)$,   $\Pi_{\frac{5}{2}}^1(p^2)$, $\Pi_{\frac{5}{2}}^2(p^2)$, $\Pi_{\frac{5}{2}}^3(p^2)$ and $\Pi_{\frac{5}{2}}^4(p^2)$.
In this article, we choose the tensor structures $g_{\mu\nu}$ and $g_{\mu\alpha}g_{\nu\beta}+g_{\mu\beta}g_{\nu\alpha}$ for analysis, thus separate the contributions of the ${\frac{3}{2}}^{\pm}$ and ${\frac{5}{2}}^{\pm}$ pentaquark states unambiguously, and tentatively assign the $P_c(4380)$ and $P_c(4450)$ to be the ${\frac{3}{2}}^-$ and ${\frac{5}{2}}^+$ pentaquark states, respectively.

The   current $J_\mu(x)$  has non-vanishing couplings  with the scattering states  $p J/\psi$, $\Lambda_c^+ \bar{D}^{*0}$, $p \chi_{c1}$ etc.
 In the following, we illustrate how to take into account  the contributions of the  intermediate   baryon-meson loops to the correlation function $\Pi_{\mu\nu}(p)$,
\begin{eqnarray}
\Pi_{\mu\nu}(p) &=&\frac{1}{ \!\not\!{p} -\widehat{M}_{-}-\Sigma^{-}_{p J/\psi}(p)-\Sigma^{-}_{\Lambda_c^+ \bar{D}^{*0}}(p)
-\Sigma^{-}_{p \chi_{c1}}(p)+\cdots} \,{\lambda^{-}_{\frac{3}{2}}}^2\,g_{\mu\nu} \nonumber\\
&&+i\gamma_5\frac{1}{ \!\not\!{p} -\widehat{M}_{+}-\Sigma^{+}_{p J/\psi}(p)-\Sigma^{+}_{\Lambda_c^+ \bar{D}^{*0}}(p)
-\Sigma^{+}_{p \chi_{c1}}(p)+\cdots} i\gamma_5\, {\lambda^{+}_{\frac{3}{2}}}^2\,g_{\mu\nu}+\cdots\, ,
\end{eqnarray}
where the $\lambda^{\pm}_{\frac{3}{2}}$ and $\widehat{M}_{\pm}$ are bare quantities to absorb the divergences in the self-energies $\Sigma^{\pm}_{p J/\psi}(p)$, $\Sigma^{\pm}_{\Lambda_c^+ \bar{D}^{*0}}(p)$, $\Sigma^{\pm}_{p \chi_{c1}}(p)$, etc.
The renormalized self-energies  contribute  a finite imaginary part to modify the dispersion relation,
\begin{eqnarray}
\Pi_{\mu\nu}(p) &=&\frac{\!\not\!{p} +M_{-}}{ p^2-M_{-}^2+i\sqrt{p^2}\Gamma_{-}(p^2)}{\lambda^{-}_{\frac{3}{2}}}^2 \, g_{\mu\nu}+\frac{\!\not\!{p} -M_{+}}{ p^2-M_{+}^2+i\sqrt{p^2}\Gamma_{+}(p^2)}{\lambda^{+}_{\frac{3}{2}}}^2\, g_{\mu\nu}\cdots \, .
 \end{eqnarray}
If we assign the $P_c(4380)$ to be the $J^P={\frac{3}{2}}^-$ pentaquark state, the width $\Gamma_{-}(p^2=M_{-}^2)=\Gamma_{P_c(4380)}=205\pm 18\pm 86\,\rm{MeV}$, which  is much smaller than the width of the $Z_c(4200)$, $\Gamma_{Z_c(4200)} = 370^{+70}_{-70}{}^{+70}_{-132}\,\rm{MeV}$.
 In Ref.\cite{Wang-Octet}, we observe that  the finite width (even as large as $400\,\rm{MeV}$) effect can be absorbed   into the pole residue $\lambda_{Z_c(4200)}$ safely, the intermediate   meson-loops cannot  affect  the mass $M_{Z_c(4200)}$ significantly,
 so the zero width approximation in  the hadronic spectral density  works. The   contributions of the  intermediate   baryon-meson loops to the correlation function $\Pi_{\mu\nu\alpha\beta}(p)$ can be studied analogously, furthermore, the width $\Gamma_{P_c(4450)}$ is much smaller than the width $\Gamma_{P_c(4380)}$.  In this article, we take the zero width approximation, which will not impair the
 predictive  ability    significantly.

Now we obtain the spectral densities at phenomenological side through the dispersion relation,
\begin{eqnarray}
\frac{{\rm Im}\Pi_{\frac{3}{2}}(s)}{\pi}&=&\!\not\!{p} \left[{\lambda^{-}_{\frac{3}{2}}}^2 \delta\left(s-M_{-}^2\right)+{\lambda^{+}_{\frac{3}{2}}}^2 \delta\left(s-M_{+}^2\right)\right] +\left[M_{-}{\lambda^{-}_{\frac{3}{2}}}^2 \delta\left(s-M_{-}^2\right)-M_{+}{\lambda^{+}_{\frac{3}{2}}}^2 \delta\left(s-M_{+}^2\right)\right]\, , \nonumber\\
&=&\!\not\!{p}\, \rho^1_{\frac{3}{2},H}(s)+\rho^0_{\frac{3}{2},H}(s) \, , \\
\frac{{\rm Im}\Pi_{\frac{5}{2}}(s)}{\pi}&=&\!\not\!{p} \left[{\lambda^{+}_{\frac{5}{2}}}^2 \delta\left(s-M_{+}^2\right)+{\lambda^{-}_{\frac{5}{2}}}^2 \delta\left(s-M_{-}^2\right)\right] +\left[M_{+}{\lambda^{+}_{\frac{5}{2}}}^2 \delta\left(s-M_{+}^2\right)-M_{-}{\lambda^{-}_{\frac{5}{2}}}^2 \delta\left(s-M_{-}^2\right)\right]\, , \nonumber\\
&=&\!\not\!{p} \,\rho^1_{\frac{5}{2},H}(s)+\rho^0_{\frac{5}{2},H}(s) \, ,
\end{eqnarray}
where the subscript $H$ denotes  the hadron side,
then we introduce the weight function $\exp\left(-\frac{s}{T^2}\right)$ to obtain the QCD sum rules at the phenomenological side (or the hadron side),
\begin{eqnarray}
\int_{4m_c^2}^{s_0}ds \left[\sqrt{s}\rho^1_{\frac{3}{2},H}(s)+\rho^0_{\frac{3}{2},H}(s)\right]\exp\left( -\frac{s}{T^2}\right)
&=&2M_{-}{\lambda^{-}_{\frac{3}{2}}}^2\exp\left( -\frac{M_{-}^2}{T^2}\right) \, ,\\
\int_{4m_c^2}^{s_0}ds \left[\sqrt{s}\rho^1_{\frac{3}{2},H}(s)-\rho^0_{\frac{3}{2},H}(s)\right]\exp\left( -\frac{s}{T^2}\right)
&=&2M_{+}{\lambda^{+}_{\frac{3}{2}}}^2\exp\left( -\frac{M_{+}^2}{T^2}\right) \, ,\\
\int_{4m_c^2}^{s_0}ds \left[\sqrt{s}\rho^1_{\frac{5}{2},H}(s)+\rho^0_{\frac{5}{2},H}(s)\right]\exp\left( -\frac{s}{T^2}\right)
&=&2M_{+}{\lambda^{+}_{\frac{5}{2}}}^2\exp\left( -\frac{M_{+}^2}{T^2}\right) \, , \\
\int_{4m_c^2}^{s_0}ds \left[\sqrt{s}\rho^1_{\frac{5}{2},H}(s)-\rho^0_{\frac{5}{2},H}(s)\right]\exp\left( -\frac{s}{T^2}\right)
&=&2M_{-}{\lambda^{-}_{\frac{5}{2}}}^2\exp\left( -\frac{M_{-}^2}{T^2}\right) \, ,
\end{eqnarray}
where the $s_0$ are the continuum threshold parameters and the $T^2$ are the Borel parameters.
We separate the  contributions  of the negative parity pentaquark states from that of the positive parity pentaquark states unambiguously.

In the following,  we briefly outline  the operator product expansion for the correlation functions $\Pi_{\mu\nu}(p)$ and $\Pi_{\mu\nu\alpha\beta}(p)$ in perturbative QCD.  We contract the $u$, $d$ and $c$ quark fields in the correlation functions
$\Pi_{\mu\nu}(p)$ and $\Pi_{\mu\nu\alpha\beta}(p)$  with Wick theorem, and obtain the results:
\begin{eqnarray}
\Pi_{\mu\nu}(p)&=&i\,\varepsilon^{ila}\varepsilon^{ijk}\varepsilon^{lmn}\varepsilon^{i^{\prime}l^{\prime}a^{\prime}}\varepsilon^{i^{\prime}j^{\prime}k^{\prime}}
\varepsilon^{l^{\prime}m^{\prime}n^{\prime}}\int d^4x e^{ip\cdot x} \nonumber\\
&&\left\{   Tr\left[\gamma_5 D_{kk^\prime}(x) \gamma_5 C U^{T}_{jj^\prime}(x)C\right] \,Tr\left[\gamma_\mu C_{nn^\prime}(x) \gamma_\nu C U^{T}_{mm^\prime}(x)C\right] C C_{a^{\prime}a}^T(-x)C \right. \nonumber\\
&&\left.-  Tr \left[\gamma_5 D_{kk^\prime}(x) \gamma_5 C U^{T}_{mj^\prime}(x)C \gamma_\mu C_{nn^\prime}(x) \gamma_\nu C U^{T}_{jm^\prime}(x)C\right] C C_{a^{\prime}a}^T(-x)C  \right\} \, ,
\end{eqnarray}
\begin{eqnarray}
\Pi_{\mu\nu\alpha\beta}(p)&=&\frac{i}{2}\,\varepsilon^{ila}\varepsilon^{ijk}\varepsilon^{lmn}\varepsilon^{i^{\prime}l^{\prime}a^{\prime}}
\varepsilon^{i^{\prime}j^{\prime}k^{\prime}}\varepsilon^{l^{\prime}m^{\prime}n^{\prime}}\int d^4x e^{ip\cdot x} \nonumber\\
&&\left\{   Tr\left[\gamma_5 D_{kk^\prime}(x) \gamma_5 C U^{T}_{jj^\prime}(x)C\right] \,Tr\left[\gamma_\mu C_{nn^\prime}(x) \gamma_\alpha C U^{T}_{mm^\prime}(x)C\right]\gamma_{\nu} C C_{a^{\prime}a}^T(-x)C\gamma_{\beta} \right. \nonumber\\
&&+Tr\left[\gamma_5 D_{kk^\prime}(x) \gamma_5 C U^{T}_{jj^\prime}(x)C\right] \,Tr\left[\gamma_\nu C_{nn^\prime}(x) \gamma_\alpha C U^{T}_{mm^\prime}(x)C\right]\gamma_{\mu} C C_{a^{\prime}a}^T(-x)C\gamma_{\beta}\nonumber\\
&&+Tr\left[\gamma_5 D_{kk^\prime}(x) \gamma_5 C U^{T}_{jj^\prime}(x)C\right] \,Tr\left[\gamma_\mu C_{nn^\prime}(x) \gamma_\beta C U^{T}_{mm^\prime}(x)C\right]\gamma_{\nu} C C_{a^{\prime}a}^T(-x)C\gamma_{\alpha}\nonumber\\
&&+Tr\left[\gamma_5 D_{kk^\prime}(x) \gamma_5 C U^{T}_{jj^\prime}(x)C\right] \,Tr\left[\gamma_\nu C_{nn^\prime}(x) \gamma_\beta C U^{T}_{mm^\prime}(x)C\right]\gamma_{\mu} C C_{a^{\prime}a}^T(-x)C\gamma_{\alpha}\nonumber\\
&&-  Tr \left[\gamma_5 D_{kk^\prime}(x) \gamma_5 C U^{T}_{mj^\prime}(x)C \gamma_\mu C_{nn^\prime}(x) \gamma_\alpha C U^{T}_{jm^\prime}(x)C\right] \gamma_\nu C C_{a^{\prime}a}^T(-x)C \gamma_\beta  \nonumber\\
&&-  Tr \left[\gamma_5 D_{kk^\prime}(x) \gamma_5 C U^{T}_{mj^\prime}(x)C \gamma_\nu C_{nn^\prime}(x) \gamma_\alpha C U^{T}_{jm^\prime}(x)C\right] \gamma_\mu C C_{a^{\prime}a}^T(-x)C \gamma_\beta  \nonumber\\
&&-  Tr \left[\gamma_5 D_{kk^\prime}(x) \gamma_5 C U^{T}_{mj^\prime}(x)C \gamma_\mu C_{nn^\prime}(x) \gamma_\beta C U^{T}_{jm^\prime}(x)C\right] \gamma_\nu C C_{a^{\prime}a}^T(-x)C \gamma_\alpha  \nonumber\\
&&\left.-  Tr \left[\gamma_5 D_{kk^\prime}(x) \gamma_5 C U^{T}_{mj^\prime}(x)C \gamma_\nu C_{nn^\prime}(x) \gamma_\beta C U^{T}_{jm^\prime}(x)C\right] \gamma_\mu C C_{a^{\prime}a}^T(-x)C \gamma_\alpha \right\} \, ,
\end{eqnarray}
where
the $U_{ij}(x)$, $D_{ij}(x)$ and $C_{ij}(x)$ are the full $u$, $d$ and $c$ quark propagators respectively ($S_{ij}(x)=U_{ij}(x),\,D_{ij}(x)$),
 \begin{eqnarray}
S_{ij}(x)&=& \frac{i\delta_{ij}\!\not\!{x}}{ 2\pi^2x^4}-\frac{\delta_{ij}\langle
\bar{q}q\rangle}{12} -\frac{\delta_{ij}x^2\langle \bar{q}g_s\sigma Gq\rangle}{192} -\frac{ig_sG^{a}_{\alpha\beta}t^a_{ij}(\!\not\!{x}
\sigma^{\alpha\beta}+\sigma^{\alpha\beta} \!\not\!{x})}{32\pi^2x^2} \nonumber\\
&&  -\frac{1}{8}\langle\bar{q}_j\sigma^{\mu\nu}q_i \rangle \sigma_{\mu\nu}+\cdots \, ,
\end{eqnarray}
\begin{eqnarray}
C_{ij}(x)&=&\frac{i}{(2\pi)^4}\int d^4k e^{-ik \cdot x} \left\{
\frac{\delta_{ij}}{\!\not\!{k}-m_c}
-\frac{g_sG^n_{\alpha\beta}t^n_{ij}}{4}\frac{\sigma^{\alpha\beta}(\!\not\!{k}+m_c)+(\!\not\!{k}+m_c)
\sigma^{\alpha\beta}}{(k^2-m_c^2)^2}\right.\nonumber\\
&&\left. -\frac{g_s^2 (t^at^b)_{ij} G^a_{\alpha\beta}G^b_{\mu\nu}(f^{\alpha\beta\mu\nu}+f^{\alpha\mu\beta\nu}+f^{\alpha\mu\nu\beta}) }{4(k^2-m_c^2)^5}+\cdots\right\} \, ,\nonumber\\
f^{\alpha\beta\mu\nu}&=&(\!\not\!{k}+m_c)\gamma^\alpha(\!\not\!{k}+m_c)\gamma^\beta(\!\not\!{k}+m_c)\gamma^\mu(\!\not\!{k}+m_c)\gamma^\nu(\!\not\!{k}+m_c)\, ,
\end{eqnarray}
and  $t^n=\frac{\lambda^n}{2}$, the $\lambda^n$ is the Gell-Mann matrix   \cite{PRT85}, then compute  the integrals both in the coordinate and momentum spaces to obtain the correlation functions $\Pi_{\mu\nu}(p)$ and $\Pi_{\mu\nu\alpha\beta}(p)$ therefore the QCD spectral densities $\rho^1_{\frac{3}{2}/\frac{5}{2},QCD}(s)$ and $\rho^0_{\frac{3}{2}/\frac{5}{2},QCD}(s)$ through the dispersion  relation.
In Eq.(37), we retain the term $\langle\bar{q}_j\sigma_{\mu\nu}q_i \rangle$  comes from the Fierz re-arrangement of the $\langle q_i \bar{q}_j\rangle$ to  absorb the gluons  emitted from both the heavy quark lines and light quark lines to form $\langle\bar{q}_j g_s G^a_{\alpha\beta} t^a_{mn}\sigma_{\mu\nu} q_i \rangle$ so as to extract the mixed condensate  $\langle\bar{q}g_s\sigma G q\rangle$.

 Once the analytical QCD spectral densities $\rho^1_{\frac{3}{2}/\frac{5}{2},QCD}(s)$ and $\rho^0_{\frac{3}{2}/\frac{5}{2},QCD}(s)$ are obtained,  we can take the
quark-hadron duality below the continuum thresholds  $s_0$ and introduce the weight function $\exp\left(-\frac{s}{T^2}\right)$ to obtain  the following QCD sum rules:
\begin{eqnarray}
2M_{-}{\lambda^{-}_{\frac{3}{2}}}^2\exp\left( -\frac{M_{-}^2}{T^2}\right)
&=& \int_{4m_c^2}^{s_0}ds \left[\sqrt{s}\rho^1_{\frac{3}{2},QCD}(s)+\rho^0_{\frac{3}{2},QCD}(s)\right]\exp\left( -\frac{s}{T^2}\right)\, ,\\
2M_{+}{\lambda^{+}_{\frac{3}{2}}}^2\exp\left( -\frac{M_{+}^2}{T^2}\right)
&=& \int_{4m_c^2}^{s_0}ds \left[\sqrt{s}\rho^1_{\frac{3}{2},QCD}(s)-\rho^0_{\frac{3}{2},QCD}(s)\right]\exp\left( -\frac{s}{T^2}\right)\, ,\\
2M_{+}{\lambda^{+}_{\frac{5}{2}}}^2\exp\left( -\frac{M_{+}^2}{T^2}\right) &=&\int_{4m_c^2}^{s_0}ds \left[\sqrt{s}\rho^1_{\frac{5}{2},QCD}(s)-\rho^0_{\frac{5}{2},QCD}(s)\right]\exp\left( -\frac{s}{T^2}\right)\, , \\
2M_{-}{\lambda^{-}_{\frac{5}{2}}}^2\exp\left( -\frac{M_{-}^2}{T^2}\right) &=&\int_{4m_c^2}^{s_0}ds \left[\sqrt{s}\rho^1_{\frac{5}{2},QCD}(s)+\rho^0_{\frac{5}{2},QCD}(s)\right]\exp\left( -\frac{s}{T^2}\right)\, ,
\end{eqnarray}
where
\begin{eqnarray}
\rho^1_{\frac{3}{2},QCD}(s)&=&\rho^1_{QCD}(s)\, ,\nonumber\\
\rho^1_{\frac{5}{2},QCD}(s)&=&2\rho^1_{QCD}(s)\, , \\
\rho^0_{\frac{3}{2},QCD}(s)&=&m_c\widetilde{\rho}^0_{QCD}(s)\, ,\nonumber\\
\rho^0_{\frac{5}{2},QCD}(s)&=&2m_c\widetilde{\rho}^0_{QCD}(s)\, ,
\end{eqnarray}
\begin{eqnarray}
\rho^1_{QCD}(s)&=&\rho_0^1(s)+\rho_3^1(s)+\rho_4^1(s)+\rho_5^1(s)+\rho_6^1(s)+\rho_8^1(s)+\rho_9^1(s)+\rho_{10}^1(s)\, , \nonumber\\
\widetilde{\rho}_{QCD}^0(s)&=&\widetilde{\rho}_0^0(s)+\widetilde{\rho}_3^0(s)+\widetilde{\rho}_4^0(s)+\widetilde{\rho}_5^0(s)+\widetilde{\rho}_6^0(s)+\widetilde{\rho}_8^0(s)
+\widetilde{\rho}_9^0(s)+\widetilde{\rho}_{10}^0(s)\, ,
\end{eqnarray}
the explicit expressions of the  QCD spectral densities $\rho_i^1(s)$ and $\widetilde{\rho}_i^0(s)$ with $i=0,\,3,\,4,\,5,\,6,\,8,\,9,\,10$ are shown in the appendix.

From Eqs.(39-44), we can see that if we set $\lambda^+_{\frac{5}{2}}=\sqrt{2}\lambda^+_{\frac{3}{2}}$ and $\lambda^-_{\frac{5}{2}}=\sqrt{2}\lambda^-_{\frac{3}{2}}$, the four QCD sum rules in Eqs.(39-42) are reduced to two QCD sum rules, the negative parity  pentaquark states have degenerate masses, and the positive parity pentaquark states also have degenerate masses. The LHCb collaboration observe that the best fit
leads to the spin-parity  assignment  $({\frac{3}{2}}^-, {\frac{5}{2}}^+)$ for the $(P_c(4380),P_c(4450))$,  other  assignments, such as
   $({\frac{3}{2}}^+, {\frac{5}{2}}^-)$ and $({\frac{5}{2}}^+, {\frac{3}{2}}^-)$,  are also acceptable \cite{LHCb-4380}. While Eqs.(39-44) indicate that the pentaquark states  with the spin-parity  $({\frac{3}{2}}^-, {\frac{5}{2}}^+)$ and $({\frac{5}{2}}^-, {\frac{3}{2}}^+)$ have degenerate masses, which
 contradicts  with the assignments $({\frac{3}{2}}^+, {\frac{5}{2}}^-)$ and $({\frac{5}{2}}^+, {\frac{3}{2}}^-)$.

In this article, we carry out the
operator product expansion to the vacuum condensates  up to dimension-10, and
assume vacuum saturation for the  higher dimension vacuum condensates, see Eqs.(35-38).
We take the truncations $n\leq 10$ and $k\leq 1$ in a consistent way,
the operators of the orders $\mathcal{O}( \alpha_s^{k})$ with $k> 1$ are  discarded.
 The condensates $\langle g_s^3 GGG\rangle$, $\langle \frac{\alpha_s GG}{\pi}\rangle^2$,
 $\langle \frac{\alpha_s GG}{\pi}\rangle\langle \bar{s} g_s \sigma Gs\rangle$ have the dimensions 6, 8, 9 respectively,  but they are   the vacuum expectations
of the operators of the order    $\mathcal{O}( \alpha_s^{3/2})$, $\mathcal{O}(\alpha_s^2)$, $\mathcal{O}( \alpha_s^{3/2})$ respectively.   Furthermore,  the numerical values of the  condensates
 $\langle \bar{q}q\rangle\langle \frac{\alpha_s}{\pi}GG\rangle$ and
$\langle \bar{q}q\rangle^2\langle \frac{\alpha_s}{\pi}GG\rangle$   are very small, and accompanied by large
denominators, and    they are neglected safely.

We differentiate   Eqs.(39-42) with respect to  $\frac{1}{T^2}$, then eliminate the
 pole residues $\lambda^{\pm}_{\frac{3}{2}(\frac{5}{2})}$ and obtain the QCD sum rules for
 the masses of the pentaquark states,
 \begin{eqnarray}
 M^2_{-} &=& \frac{\int_{4m_c^2}^{s_0}ds \,s\,\left[\sqrt{s}\rho^1_{QCD}(s)+m_c\widetilde{\rho}^0_{QCD}(s)\right]\exp\left( -\frac{s}{T^2}\right)}{\int_{4m_c^2}^{s_0}ds \left[\sqrt{s}\rho_{QCD}^1(s)+m_c\widetilde{\rho}^0_{QCD}(s)\right]\exp\left( -\frac{s}{T^2}\right)}\, ,\\
M^2_{+} &=& \frac{\int_{4m_c^2}^{s_0}ds \,s\,\left[\sqrt{s}\rho^1_{QCD}(s)-m_c\widetilde{\rho}^0_{QCD}(s)\right]\exp\left( -\frac{s}{T^2}\right)}{\int_{4m_c^2}^{s_0}ds \left[\sqrt{s}\rho_{QCD}^1(s)-m_c\widetilde{\rho}^0_{QCD}(s)\right]\exp\left( -\frac{s}{T^2}\right)}\, ,
\end{eqnarray}
where the  $M_{-}$ ($M_{+}$) are the masses of the $J^P={\frac{3}{2}}^-,\,{\frac{5}{2}}^-$ (${\frac{3}{2}}^+,\,{\frac{5}{2}}^+$) pentaquark states.
Once the masses $M_{\pm}$ are obtained, we can take them as input parameters and obtain the pole residues from the QCD sum rules in Eqs.(39-42), the relations $\lambda^+_{\frac{5}{2}}=\sqrt{2}\lambda^+_{\frac{3}{2}}$ and $\lambda^-_{\frac{5}{2}}=\sqrt{2}\lambda^-_{\frac{3}{2}}$ hold.

\section{Numerical results and discussions}
We take the vacuum condensates to be  the standard values
$\langle\bar{q}q \rangle=-(0.24\pm 0.01\, \rm{GeV})^3$,
$\langle\bar{q}g_s\sigma G q \rangle=m_0^2\langle \bar{q}q \rangle$,
$m_0^2=(0.8 \pm 0.1)\,\rm{GeV}^2$, $\langle \frac{\alpha_s
GG}{\pi}\rangle=(0.33\,\rm{GeV})^4 $    at the energy scale  $\mu=1\, \rm{GeV}$
\cite{SVZ79,PRT85}.
The quark condensates and mixed quark condensates  evolve with the   renormalization group equation,
$\langle\bar{q}q \rangle(\mu)=\langle\bar{q}q \rangle(Q)\left[\frac{\alpha_{s}(Q)}{\alpha_{s}(\mu)}\right]^{\frac{4}{9}}$ and
 $\langle\bar{q}g_s \sigma Gq \rangle(\mu)=\langle\bar{q}g_s \sigma Gq \rangle(Q)\left[\frac{\alpha_{s}(Q)}{\alpha_{s}(\mu)}\right]^{\frac{2}{27}}$.
In the article, we take the $\overline{MS}$ mass $m_{c}(m_c)=(1.275\pm0.025)\,\rm{GeV}$
 from the Particle Data Group \cite{PDG}, and take into account
the energy-scale dependence of  the $\overline{MS}$ mass from the renormalization group equation,
\begin{eqnarray}
m_c(\mu)&=&m_c(m_c)\left[\frac{\alpha_{s}(\mu)}{\alpha_{s}(m_c)}\right]^{\frac{12}{25}} \, ,\nonumber\\
\alpha_s(\mu)&=&\frac{1}{b_0t}\left[1-\frac{b_1}{b_0^2}\frac{\log t}{t} +\frac{b_1^2(\log^2{t}-\log{t}-1)+b_0b_2}{b_0^4t^2}\right]\, ,
\end{eqnarray}
  where $t=\log \frac{\mu^2}{\Lambda^2}$, $b_0=\frac{33-2n_f}{12\pi}$, $b_1=\frac{153-19n_f}{24\pi^2}$, $b_2=\frac{2857-\frac{5033}{9}n_f+\frac{325}{27}n_f^2}{128\pi^3}$,  $\Lambda=213\,\rm{MeV}$, $296\,\rm{MeV}$  and  $339\,\rm{MeV}$ for the flavors  $n_f=5$, $4$ and $3$, respectively  \cite{PDG}.

In Refs.\cite{Wang-Tetraquark-DW, Wang-molecule,Wang-Octet,WangHuang-PRD,Wang-Tetraquark-DW-2}, we study the acceptable energy scales of the QCD spectral densities  for the hidden  charm (bottom) tetraquark states and molecular (and molecule-like) states in the QCD sum rules in details for the first time,  and suggest a  formula $\mu=\sqrt{M^2_{X/Y/Z}-(2{\mathbb{M}}_Q)^2}$ to determine  the energy scales, where the $X$, $Y$, $Z$ denote the four-quark systems, and the ${\mathbb{M}}_Q$ denotes the effective heavy quark masses.
The effective mass ${\mathbb{M}}_c=1.8\,\rm{GeV}$ is the optimal value for  the diquark-antidiquark type tetraquark states \cite{Wang-Tetraquark-DW, WangHuang-PRD,Wang-Tetraquark-DW-2}.

In this article, we use the diquark-diquark-antiquark model to construct the currents to interpolate the hidden-charm pentaquark states, there also exists  a $\bar{c}c$ quark pair.
The hidden charm (or bottom) five-quark systems  $qq_1q_2Q\bar{Q}$ could be described
by a double-well potential, just like the four-quark systems $qq^{\prime}Q\bar{Q}$, see Eqs.(3-8) and related discussions in the introduction.
The  heavy five-quark states  are also characterized by the effective heavy quark masses ${\mathbb{M}}_Q$ and
the virtuality $V=\sqrt{M^2_{P_c}-(2{\mathbb{M}}_Q)^2}$.   The  QCD sum rules have three typical energy scales $\mu^2$, $T^2$, $V^2$, we can also
 take the energy  scale, $ \mu^2=V^2={\mathcal{O}}(T^2)$ \cite{Wang-Tetraquark-DW,Wang-Tetraquark-DW-2}. In this article,
 we can take the analogous formula,
 \begin{eqnarray}
 \mu&=&\sqrt{M_{P_c}^2-(2{\mathbb{M}}_c)^2}\, ,
   \end{eqnarray}
   with  the value ${\mathbb{M}}_c=1.8\,\rm{GeV}$ to determine the energy scales of the QCD spectral densities \cite{Wang-Tetraquark-DW,Wang-Tetraquark-DW-2}, and obtain the values $\mu=2.5\,\rm{GeV}$ and $\mu=2.6\,\rm{GeV}$ for the hidden charm pentaquark states $P_c(4380)$
 and $P_c(4450)$, respectively. The energy scale formula can be rewritten as
 \begin{eqnarray}
 M_{P_c}^2&=&(2{\mathbb{M}}_c)^2+\mu^2\, .
 \end{eqnarray}

In this article, we choose the  Borel parameters $T^2$ and continuum threshold
parameters $s_0$  to satisfy the  following criteria:

$\bf{1_\cdot}$ Pole dominance at the phenomenological side;

$\bf{2_\cdot}$ Convergence of the operator product expansion;

$\bf{3_\cdot}$ Appearance of the Borel platforms;

$\bf{4_\cdot}$ Satisfying the energy scale formula.

In the QCD sum rules for the multiquark states, it is difficult to satisfy  the criteria $\bf{1}$ and $\bf{2}$. In previous work \cite{Wang-Tetraquark-DW,WangHuang-PRD}, we observed that the pole contributions can be taken as large as $(50-70)\%$ in the QCD sum rules for the diquark-antidiquark type tetraquark states $qq^{\prime}Q\bar{Q}$ ($X,Y,Z$), if the QCD spectral densities obey  the energy scale formula $\mu=\sqrt{M_{X/Y/Z}^2-(2{\mathbb{M}}_Q)^2}$. The operator product expansion converges
more slowly  in the QCD sum rules for the pentaquark states $qq_1q_2Q\bar{Q}$ compared  to  that for the  tetraquark states $q\bar{q}^{\prime}Q\bar{Q}$, so in this article, we choose smaller pole contributions, about $(50\pm10)\%$. For the  tetraquark states $q\bar{q}^{\prime}Q\bar{Q}$ \cite{Wang-Tetraquark-DW,WangHuang-PRD}, the   Borel platforms appear as the minimum values, and the platforms are very flat, but the Borel windows are small, $T^2_{max}-T^2_{min}=0.4\,\rm{GeV}^2$, where the $max$ and $min$ denote the maximum and minimum values, respectively. For the three-quark baryons $qq^{\prime}Q$, $qQQ^\prime$, $QQ^{\prime}Q^{\prime\prime}$  \cite{WangLambda,WangHbaryon},  the Borel platforms do not appear as the minimum values, the predicted masses increase slowly with the increase of the Borel parameter, we determine the Borel windows by the criteria $\bf{1}$ and $\bf{2}$, the platforms are not very flat. In this article, we also choose small Borel windows $T^2_{max}-T^2_{min}=0.4\,\rm{GeV}^2$, just like in the case of the tetraquark states,  and obtain the platforms  by requiring the uncertainties $\frac{\delta M_{P_c}}{M_{P_c}} $ induced by the Borel parameters are about $1\%$.

Now we  search for the optimal  Borel parameters $T^2$ and continuum threshold parameters $s_0$ according to  the four criteria. The resulting Borel parameters, continuum threshold parameters, energy scales, pole contributions are shown explicitly in Table 1. Furthermore, the contributions of the vacuum condensates of dimension 10 are less than $5\%$, the operator product expansion is convergent.
    So the four criteria of the QCD sum rules are satisfied, we expect to obtain reasonable predictions. From   Table 1, we can see that the values
$\sqrt{s_0}= M_{P_c({\frac{3}{2}}^-,{\frac{5}{2}}^+)}+(0.6-0.8)\,\rm{GeV}$ (\, or $s_0^{P_c({\frac{3}{2}}^-)}=(26\pm1)\,\rm{GeV}^2$, $s_0^{P_c({\frac{5}{2}}^+)}=(27\pm1)\,\rm{GeV}^2$\,) can lead to satisfactory results.
\begin{figure}
 \centering
 \includegraphics[totalheight=6cm,width=9cm]{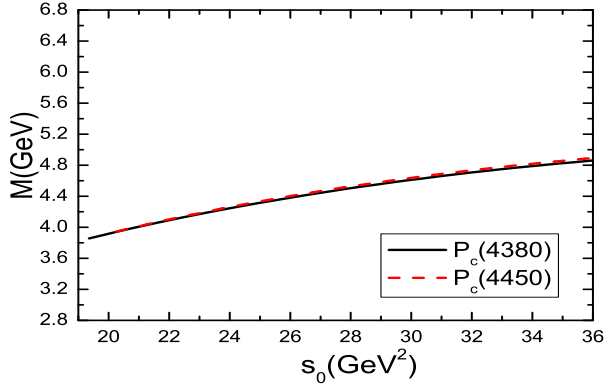}
         \caption{ The masses  of the pentaquark states  with variations of the threshold  parameters $s_0$.  }
\end{figure}

In Fig.1, we plot the  predicted masses with variation of the
threshold parameters $s_0$, where we assign the $P_c(4380)$ and $P_c(4450)$ to be the ${\frac{3}{2}}^-$ and ${\frac{5}{2}}^+$ pentaquark states, respectively.
 From the figure, we can see that the
predicted masses increase slowly with (or are not  sensitive to) the threshold
parameters $s_0$  for central values of other parameters.

In Refs.\cite{WangLambda,WangHbaryon},  we  study the  $J^P={1\over 2}^{\pm}$ and ${3\over 2}^{\pm}$ heavy, doubly-heavy and triply-heavy baryon states  systematically   with the QCD sum rules by subtracting the contributions from the corresponding $J^P={1\over 2}^{\mp}$ and ${3\over 2}^{\mp}$  heavy, doubly-heavy and triply-heavy baryon states, the continuum threshold parameters $\sqrt{s_0}=M_{\rm{gr}}+ (0.6-0.8)\,\rm{GeV}$ work well, where subscript $\rm{gr}$ denotes the ground states.
In the present case, the hidden charm pentaquark states carry a baryon number of one, i.e. they are  doubly-heavy baryons.
So the threshold parameters $\sqrt{s_0}= M_{P_c({\frac{3}{2}}^-,{\frac{5}{2}}^+)}+(0.6-0.8)\,\rm{GeV}$ make sense.
One may worry that there exist some contaminations from the higher resonances, the upper bounds of the factors $\exp\left(-\frac{s_0}{T^2} \right)$ are about $0.0007$ and $0.0004$ in the QCD sum rules for the $P_c(4380)$ and $P_c(4450)$, respectively,  if we take the largest values of the continuum  threshold parameters, so the contaminations are greatly suppressed and can be neglected safely.

\begin{table}
\begin{center}
\begin{tabular}{|c|c|c|c|c|c|c|c|}\hline\hline
                        &$T^2(\rm{GeV}^2)$  &$\sqrt{s_0}(\rm{GeV})$ &$\mu(\rm{GeV})$ &pole         &$M_{P_c}(\rm{GeV})$  &$\lambda_{P_c}(\rm{GeV}^6)$ \\ \hline
$P_c({\frac{3}{2}}^-)$  &$3.3-3.7$          &$5.10\pm0.10$          &$2.5$           &$(40-61)\%$  &$4.38\pm0.13$        &$(1.55\pm0.28)\times10^{-3}$  \\ \hline
$P_c({\frac{5}{2}}^+)$  &$3.1-3.5$          &$5.15\pm0.10$          &$2.6$           &$(40-63)\%$  &$4.44\pm0.14$        &$(0.84\pm0.17)\times10^{-3}$  \\ \hline
 \hline
\end{tabular}
\end{center}
\caption{ The Borel parameters, continuum threshold parameters, energy scales,  pole contributions, masses and pole residues of the pentaquark states. }
\end{table}

We take into account  all uncertainties  of the input   parameters,
and obtain the values of the masses and pole residues of
 the ${3\over 2}^-$ and ${5\over 2}^+$ hidden-charm pentaquark states, which are shown in Figs.2-3 and Table 1.
 The QCD sum rules in Eqs.(39-42) and Eqs.(46-47) indicate that the pentaquark states  with the spin-parity  $({\frac{3}{2}}^-, {\frac{5}{2}}^+)$ and $({\frac{5}{2}}^-, {\frac{3}{2}}^+)$ have degenerate masses, and $\lambda^+_{\frac{5}{2}}=\sqrt{2}\lambda^+_{\frac{3}{2}}$ and $\lambda^-_{\frac{5}{2}}=\sqrt{2}\lambda^-_{\frac{3}{2}}$.
 Naively, we expect that additional one unit spin or P-wave can lead to larger masses, so $M_{{\frac{5}{2}}^+}>M_{{\frac{3}{2}}^-} $, while the relation  $M_{{\frac{3}{2}}^+}> M_{{\frac{5}{2}}^-} $ needs detailed and refined analysis to obtain the answer "yes" or "no". It is sensible to  assign the $P_c(4380)$ and $P_c(4450)$ to be the ${\frac{3}{2}}^-$ and ${\frac{5}{2}}^+$ pentaquark states, respectively. However, the assignment $({\frac{5}{2}}^-, {\frac{3}{2}}^+)$ of the $(P_c(4380),P_c(4450))$ is not excluded.

 \begin{figure}
 \centering
 \includegraphics[totalheight=5cm,width=7cm]{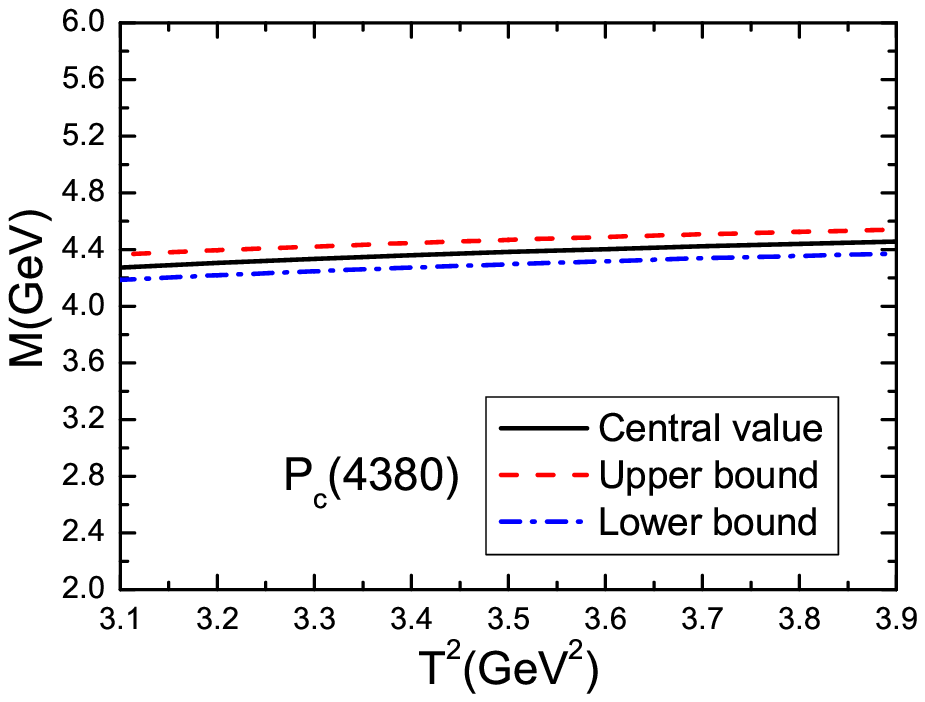}
 \includegraphics[totalheight=5cm,width=7cm]{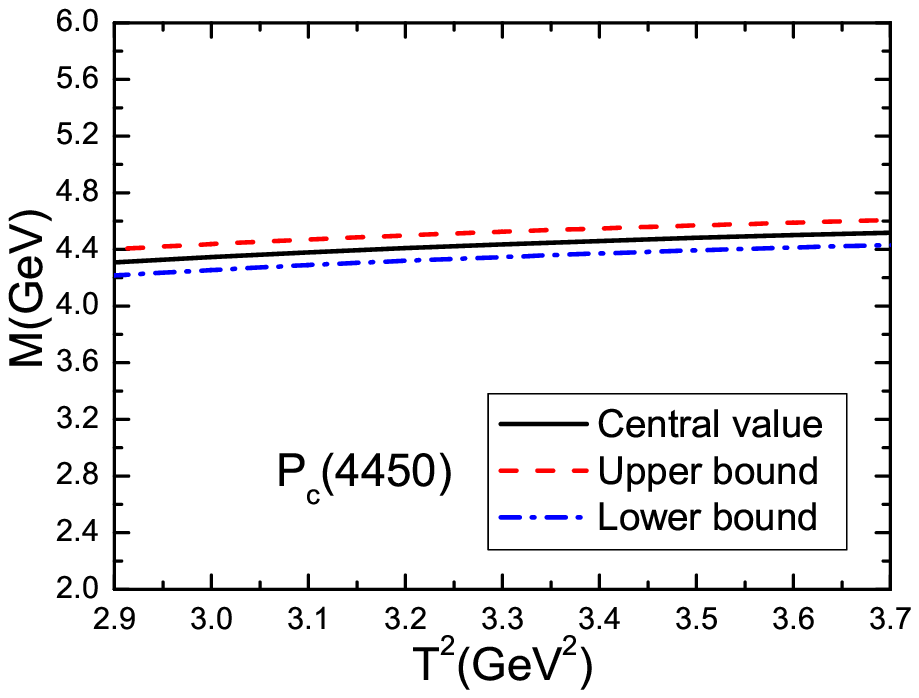}
        \caption{ The masses  of the pentaquark states  with variations of the Borel parameters $T^2$.  }
\end{figure}

\begin{figure}
 \centering
 \includegraphics[totalheight=5cm,width=7cm]{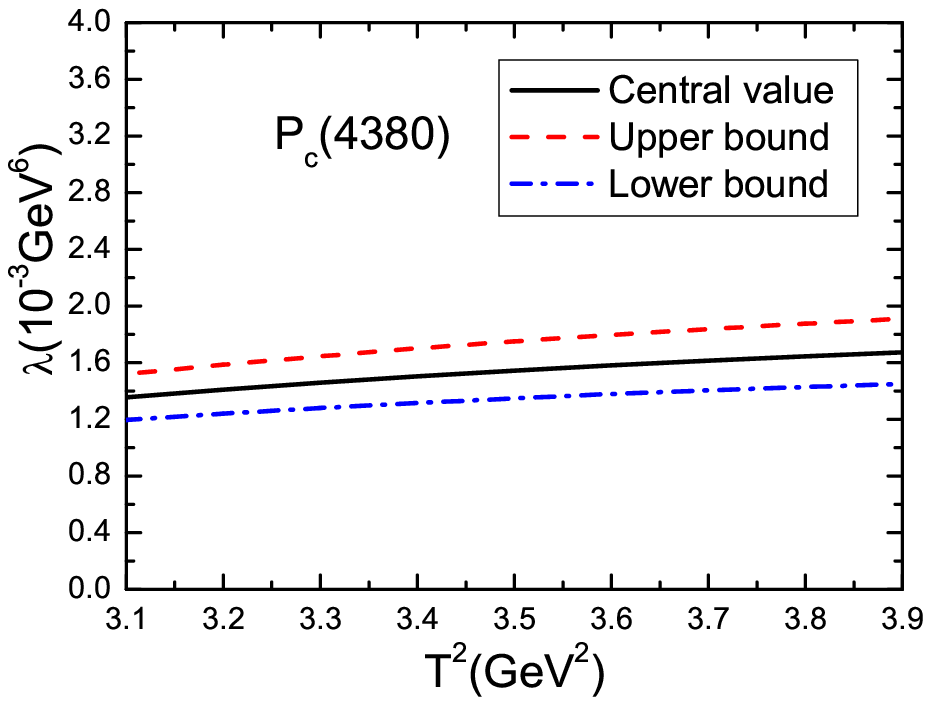}
 \includegraphics[totalheight=5cm,width=7cm]{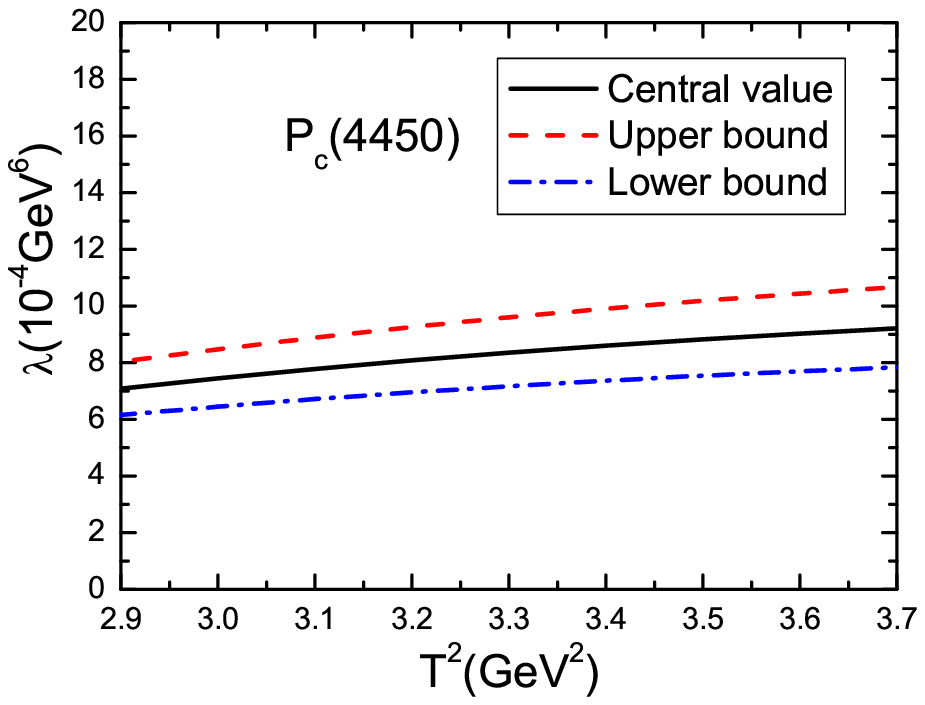}
        \caption{ The pole residues  of the pentaquark states  with variations of the Borel parameters $T^2$.  }
\end{figure}

 From Table 1, we can see that the present predictions $M_{P_c(4380)}=4.38\pm0.13\,\rm{GeV}$  and $M_{P_c(4450)}=4.44\pm0.14\,\rm{GeV}$ are in good agreement with the
experimental data of the LHCb collaboration,      $M_{P_c(4380)}=4380\pm 8\pm 29\,\rm{MeV}$  and $M_{P_c(4450)}=4449.8\pm 1.7\pm 2.5\,\rm{MeV}$ \cite{LHCb-4380}.
  The present predictions  support assigning  the $P_c(4380)$ and $P_c(4450)$ to be the  ${\frac{3}{2}}^-$ and ${\frac{5}{2}}^+$ hidden charm pentaquark states, respectively, which are consistent with the assignments that the $P_c(4380)$ and $P_c(4450)$ are diquark-diquark-antiquark type pentaquark states \cite{di-di-anti-penta} or the diquark-triquark  type  pentaquark states \cite{di-tri-penta}.

In this article, we  take the energy scale formula $ \mu=\sqrt{M_{P_c}^2-(2{\mathbb{M}}_c)^2}$ to determine the energy scales of the QCD spectral densities. The pole contributions are about $(40-60)\%$, and the  contributions of the vacuum condensates of dimension 10 are less than $5\%$, the two criteria
(pole dominance at the phenomenological side and convergence of the operator product expansion) of the conventional QCD sum rules can be satisfied, so we expect to make reasonable predictions. In subsequent works, we extend the present work to study the ${\frac{1}{2}}^\pm$ and ${\frac{3}{2}}^\pm$ hidden-charm pentaquark states in a systematic way \cite{WangPentaquark}, where the energy scale formula $ \mu=\sqrt{M_{P_c}^2-(2{\mathbb{M}}_c)^2}$ serves as an additional constraint on the predicted masses.  The typical energy scales, which  characterize the five-quark systems  $q_1q_2q_3c\bar{c}$  and serve as the optimal energy scales of the QCD spectral densities, are not independent of the masses of the five-quark systems  $q_1q_2q_3c\bar{c}$. All the predictions can be confronted to the experimental data in the future.

The diquark-diquark-antiquark type current with special quantum numbers couples potentially  to  special pentaquark states  according to the tensor analysis in Eqs.(21-22) and Eqs.(25-26). The current can be re-arranged both in the color and Dirac-spinor  spaces, and changed  to a current as a special superposition of  the color singlet  baryon-meson type currents.   The baryon-meson type currents couple potentially  to the baryon-meson pairs. The
diquark-diquark-antiquark type pentaquark state can be taken as a special superposition of a series of  baryon-meson pairs, and embodies  the net effects. The decays to its components (baryon-meson pairs) are Okubo-Zweig-Iizuka super-allowed, but the re-arrangements in the color-space are non-trivial \cite{Nielsen3900}.

In the following, we perform Fierz re-arrangement  to the currents $J_{\mu}$ and $J_{\mu\nu}$ both in the color and Dirac-spinor  spaces to   obtain the results,
\begin{eqnarray}
J_\mu &=&\frac{1}{4}\mathcal{S}c\,\bar{c}\gamma_\mu u+\frac{1}{4}\mathcal{S}u\,\bar{c}\gamma_\mu c-\frac{1}{4}\mathcal{S}\gamma_5 c\,\bar{c}\gamma_\mu \gamma_5 u-\frac{1}{4}\mathcal{S}\gamma_5 u\,\bar{c}\gamma_\mu \gamma_5c -\frac{i}{4}\mathcal{S}\gamma_\mu\gamma_5 c\,\bar{c} i\gamma_5 u-\frac{i}{4}\mathcal{S}\gamma_\mu \gamma_5 u\,\bar{c}i \gamma_5c\nonumber\\
&&-\frac{1}{4}\mathcal{S}\gamma_\mu c\,\bar{c} u-\frac{1}{4}\mathcal{S}\gamma_\mu u\,\bar{c} c -\frac{i}{4}\mathcal{S}\sigma_{\lambda\mu} c\,\bar{c}\gamma^\lambda u
-\frac{i}{4}\mathcal{S}\sigma_{\lambda\mu} u\,\bar{c}\gamma^\lambda c+\frac{i}{4}\mathcal{S}\sigma_{\lambda\mu}\gamma_5 c\,\bar{c}\gamma^\lambda \gamma_5u
\nonumber\\
&&+\frac{i}{4}\mathcal{S}\sigma_{\lambda\mu}\gamma_5 u\,\bar{c}\gamma^\lambda \gamma_5c +\frac{1}{8}\mathcal{S}\sigma_{\lambda\tau} \gamma_\mu c\,\bar{c}\sigma^{\lambda\tau} u+\frac{1}{8}\mathcal{S}\sigma_{\lambda\tau} \gamma_\mu u\,\bar{c}\sigma^{\lambda\tau} c\, ,
\end{eqnarray}
\begin{eqnarray}
  \widehat{J}_{\mu\nu}
  &=&\frac{1}{2\sqrt{2}}  \mathcal{S}\left(g_{\nu\lambda}\gamma_\mu+g_{\mu\lambda}\gamma_\nu\right)c\,\bar{c}\gamma^\lambda u
  +\frac{1}{2\sqrt{2}}\mathcal{S}\left(g_{\nu\lambda}\gamma_\mu+g_{\mu\lambda}\gamma_\nu\right)u\,\bar{c}\gamma^\lambda c    \nonumber\\
  &&-\frac{1}{2\sqrt{2}}\mathcal{S}\left(g_{\nu\lambda}\gamma_\mu+g_{\mu\lambda}\gamma_\nu\right)\gamma_5 c\,\bar{c}\gamma^\lambda \gamma_5 u
  -\frac{1}{2\sqrt{2}} \mathcal{S}\left(g_{\nu\lambda}\gamma_\mu+g_{\mu\lambda}\gamma_\nu\right)\gamma_5 u\,\bar{c}\gamma^\lambda \gamma_5 c\nonumber\\
  && +\frac{1}{8\sqrt{2}}\mathcal{S}\left(\gamma_\mu \sigma_{\lambda\tau}\gamma_\nu+\gamma_\nu \sigma_{\lambda\tau}\gamma_\mu\right)c\,\bar{c}\sigma^{\lambda\tau} u+\frac{1}{8\sqrt{2}}\mathcal{S}\left(\gamma_\mu \sigma_{\lambda\tau}\gamma_\nu+\gamma_\nu \sigma_{\lambda\tau}\gamma_\mu\right)u\,\bar{c}\sigma^{\lambda\tau} c \, ,
  \end{eqnarray}
  where we take  the replacement $J_{\mu\nu}\to \widehat{J}_{\mu\nu} $,
  \begin{eqnarray}
 J_{\mu\nu} &\to & \widehat{J}_{\mu\nu}\, ,\nonumber \\
 &=&\frac{1}{\sqrt{2}}\varepsilon^{ila} \varepsilon^{ijk}\varepsilon^{lmn}  u^T_j C\gamma_5 d_k\left[u^T_m C\gamma_\mu c_n\, \gamma_{\nu}C\bar{c}^{T}_{a}+u^T_m C\gamma_\nu c_n\, \gamma_{\mu}C\bar{c}^{T}_{a}-\frac{1}{2}g_{\mu\nu}u^T_m C\gamma_\lambda c_n\, \gamma^{\lambda}C\bar{c}^{T}_{a}\right] \, , \nonumber \\
 \end{eqnarray}
  to subtract the contribution of the spin-$\frac{1}{2}$ pentaquark state, and use   the notations  $\mathcal{S}\Gamma c=\varepsilon^{ijk}u^T_i C \gamma_5 d_j \Gamma c_k$ and $\mathcal{S}\Gamma u=\varepsilon^{ijk}u^T_i C \gamma_5 d_j \Gamma u_k$ for simplicity, here the $\Gamma$ denotes the Dirac matrixes.

  The components $\mathcal{S}(x)\Gamma c(x) \bar{c}(x)\Gamma^{\prime}u(x)$ and $\mathcal{S}(x)\Gamma u(x) \bar{c}(x)\Gamma^{\prime}c(x)$ couple potentially to the baryon-meson pairs. The revelent thresholds are $M_{J/\psi p}=4.035\,\rm{GeV}$, $M_{\eta_c p}=3.922\,\rm{GeV}$, $M_{\eta_c N(1440)}=4.414\,\rm{GeV}$,  $M_{\chi_{c0} p}=4.353\,\rm{GeV}$,  $M_{\Lambda_c^+ \bar{D}^0}=4.151\,\rm{GeV}$, $M_{\Lambda_c^+ \bar{D}^{*0}}=4.293\,\rm{GeV}$,
   $M_{h_{c} p}=4.463\,\rm{GeV}$, $M_{\chi_{c1} p}=4.449\,\rm{GeV}$, and $M_{\Lambda_c^+(2595) \bar{D}^0}=4.457\,\rm{GeV}$ \cite{PDG}. After taking into account the currents-hadrons duality, we obtain the Okubo-Zweig-Iizuka super-allowed decays,
\begin{eqnarray}
P_c(4380) &\to& p J/\psi  \, , \, \Lambda_c^+ \bar{D}^{*0}\, , \, p\eta_c \, , \,\Lambda_c^+ \bar{D}^{0}\, , \, p\chi_{c0} \, , \\
P_c(4450) &\to& p J/\psi  \, , \, \Lambda_c^+ \bar{D}^{*0}\, , \, p\eta_c\, , \,\Lambda_c^+ \bar{D}^{0}\, , \, N(1440) \eta_c\, .
\end{eqnarray}
We can search for the $P_c(4380)$ and $P_c(4450)$ in the $\Lambda_c^+ \bar{D}^{*0}$, $p\eta_c$, $\Lambda_c^+ \bar{D}^{0}$, $p\chi_{c0}$, $N(1440) \eta_c$ mass distributions in the future, which may shed light on the nature of those pentaquark states.

\section{Conclusion}
In this article, we construct the diquark-diquark-antiquark type interpolating currents,  and  study the masses and pole residues of the  ${\frac{3}{2}}^-$ and ${\frac{5}{2}}^+$ hidden-charm pentaquark states   in details with the QCD sum rules by calculating the contributions of the vacuum condensates up to dimension-10 in the operator product expansion. In calculations,  we use the  formula $\mu=\sqrt{M^2_{P_c}-(2{\mathbb{M}}_c)^2}$  to determine  the energy scales of the QCD spectral densities.  The present predictions favor assigning  the $P_c(4380)$ and $P_c(4450)$ to be the  ${\frac{3}{2}}^-$ and ${\frac{5}{2}}^+$ pentaquark states, respectively. The  pole residues can be taken as   basic input parameters to study relevant processes of the pentaquark   states with the three-point QCD sum rules.

\section*{Acknowledgements}
This  work is supported by National Natural Science Foundation,
Grant Numbers 11375063, and Natural Science Foundation of Hebei province, Grant Number A2014502017.

\section*{Appendix}
The QCD spectral densities $\rho^1_{i}(s)$ and $\widetilde{\rho}^0_{i}(s)$ with $i=0,\,3,\,4,\,5,\,6,\,8,\,9,\,10$ of the pentaquark states,

\begin{eqnarray}
\rho^1_{0}(s)&=&\frac{1}{491520\pi^8}\int dydz \, yz(1-y-z)^4\left(s-\overline{m}_c^2\right)^4\left(7s-2\overline{m}_c^2 \right) \, , \nonumber\\
\widetilde{\rho}^0_{0}(s)&=&\frac{1}{983040\pi^8}\int dydz \, (y+z)(1-y-z)^4\left(s-\overline{m}_c^2\right)^4\left(6s-\overline{m}_c^2 \right) \, ,
\end{eqnarray}

\begin{eqnarray}
\rho^1_{3}(s)&=&- \frac{m_c\langle \bar{q}q\rangle}{3072\pi^6}\int dydz \, (y+z)(1-y-z)^2\left(s-\overline{m}_c^2\right)^3  \, ,\nonumber \\
\widetilde{\rho}^0_{3}(s)&=&- \frac{m_c\langle \bar{q}q\rangle}{1536\pi^6}\int dydz \, (1-y-z)^2\left(s-\overline{m}_c^2\right)^3  \, ,
\end{eqnarray}
\begin{eqnarray}
\rho^1_{4}(s)&=&-\frac{m_c^2}{73728\pi^6} \langle\frac{\alpha_s GG}{\pi}\rangle\int dydz \left( \frac{z}{y^2}+\frac{y}{z^2}\right)(1-y-z)^4 \left(s-\overline{m}_c^2\right)\left(2s-\overline{m}_c^2\right)  \nonumber\\
&&-\frac{19}{7077888\pi^6} \langle\frac{\alpha_s GG}{\pi}\rangle\int dydz \left( y+z\right)(1-y-z)^3 \left(s-\overline{m}_c^2\right)^2\left(7s-4\overline{m}_c^2\right)  \nonumber\\
&&+\frac{13}{393216\pi^6} \langle\frac{\alpha_s GG}{\pi}\rangle\int dydz \, yz(1-y-z)^2 \left(s-\overline{m}_c^2\right)^2\left(5s-2\overline{m}_c^2\right) \, , \nonumber\\
\widetilde{\rho}^0_{4}(s)&=&-\frac{m_c^2}{294912\pi^6} \langle\frac{\alpha_s GG}{\pi}\rangle\int dydz \left( \frac{1}{y^2}+\frac{1}{z^2}+\frac{y}{z^3}+\frac{z}{y^3}\right)(1-y-z)^4 \left(s-\overline{m}_c^2\right)\left(3s-\overline{m}_c^2\right)  \nonumber\\
&&+\frac{1}{294912\pi^6} \langle\frac{\alpha_s GG}{\pi}\rangle\int dydz \left( \frac{y}{z^2}+\frac{z}{y^2}\right)(1-y-z)^4 \left(s-\overline{m}_c^2\right)^2\left(4s-\overline{m}_c^2\right)  \nonumber\\
&&-\frac{19}{1179648\pi^6} \langle\frac{\alpha_s GG}{\pi}\rangle\int dydz \,(1-y-z)^3 \left(s-\overline{m}_c^2\right)^2\left(2s-\overline{m}_c^2\right)  \nonumber\\
&&+\frac{13}{786432\pi^6} \langle\frac{\alpha_s GG}{\pi}\rangle\int dydz \,(y+z)(1-y-z)^2 \left(s-\overline{m}_c^2\right)^2\left(4s-\overline{m}_c^2\right)  \, ,
\end{eqnarray}

\begin{eqnarray}
\rho^1_{5}(s)&=& \frac{m_c\langle \bar{q}g_s\sigma Gq\rangle}{2048\pi^6}\int dydz  \, (y+z)(1-y-z) \left(s-\overline{m}_c^2 \right)^2 \nonumber\\
&&+\frac{m_c\langle \bar{q}g_s\sigma Gq\rangle}{65536\pi^6}\int dydz  \, \left(\frac{y}{z}+\frac{z}{y}\right)(1-y-z)^2 \left(s-\overline{m}_c^2 \right)^2 \nonumber\\
&&-\frac{m_c\langle \bar{q}g_s\sigma Gq\rangle}{98304\pi^6}\int dydz  \, \left(\frac{y}{z}+\frac{z}{y}\right)(1-y-z)^3 \left(s-\overline{m}_c^2 \right)^2 \nonumber\\
&&+\frac{3m_c\langle \bar{q}g_s\sigma Gq\rangle}{32768\pi^6}\int dydz  \, \left(y+z\right)(1-y-z) \left(s-\overline{m}_c^2 \right)^2 \, ,\nonumber\\
\widetilde{\rho}^0_{5}(s)&=& \frac{m_c\langle \bar{q}g_s\sigma Gq\rangle}{1024\pi^6}\int dydz  \, (1-y-z) \left(s-\overline{m}_c^2 \right)^2 \nonumber\\
&&+\frac{m_c\langle \bar{q}g_s\sigma Gq\rangle}{65536\pi^6}\int dydz  \, \left(\frac{1}{y}+\frac{1}{z}\right)(1-y-z)^2 \left(s-\overline{m}_c^2 \right)^2 \nonumber\\
&&-\frac{m_c\langle \bar{q}g_s\sigma Gq\rangle}{98304\pi^6}\int dydz  \, \left(\frac{1}{y}+\frac{1}{z}\right)(1-y-z)^3 \left(s-\overline{m}_c^2 \right)^2 \nonumber\\
&&+\frac{3m_c\langle \bar{q}g_s\sigma Gq\rangle}{16384\pi^6}\int dydz  \,  (1-y-z) \left(s-\overline{m}_c^2 \right)^2 \, ,
\end{eqnarray}

\begin{eqnarray}
\rho^1_{6}(s)&=&\frac{\langle\bar{q}q\rangle^2}{96\pi^4}\int dydz \,  yz(1-y-z)\left(s-\overline{m}_c^2 \right)\left(2s-\overline{m}_c^2 \right)\,,\nonumber\\
\widetilde{\rho}^0_{6}(s)&=&\frac{\langle\bar{q}q\rangle^2}{384\pi^4}\int dydz \,  (y+z)(1-y-z)\left(s-\overline{m}_c^2 \right)\left(3s-\overline{m}_c^2 \right)\, ,
\end{eqnarray}

\begin{eqnarray}
\rho^1_8(s)&=&-\frac{35\langle\bar{q}q\rangle\langle\bar{q}g_s\sigma Gq\rangle}{6144\pi^4}\int dydz \,yz \left(3s-2\overline{m}_c^2\right)\nonumber\\
&&-\frac{\langle\bar{q}q\rangle\langle\bar{q}g_s\sigma Gq\rangle}{12288\pi^4}\int dydz \,(y+z)(1-y-z) \left(5s-4\overline{m}_c^2\right)\, ,\nonumber\\
\widetilde{\rho}^0_8(s)&=&-\frac{35\langle\bar{q}q\rangle\langle\bar{q}g_s\sigma Gq\rangle}{12288\pi^4}\int dydz \,(y+z) \left(2s-\overline{m}_c^2\right)\nonumber\\
&&-\frac{\langle\bar{q}q\rangle\langle\bar{q}g_s\sigma Gq\rangle}{6144\pi^4}\int dydz \, (1-y-z) \left(4s-3\overline{m}_c^2\right)\, ,
\end{eqnarray}

\begin{eqnarray}
\rho^1_{9}(s)&=&-\frac{m_c\langle\bar{q}q\rangle^3}{144\pi^2}\int_{y_i}^{y_f} dy   \, , \nonumber\\
\widetilde{\rho}^0_{9}(s)&=&-\frac{m_c\langle\bar{q}q\rangle^3}{72\pi^2}\int_{y_i}^{y_f} dy   \,,
\end{eqnarray}

\begin{eqnarray}
\rho^1_{10}(s)&=&\frac{19\langle\bar{q}g_s\sigma Gq\rangle^2}{24576\pi^4}\int_{y_i}^{y_f} dy \, y(1-y)\left[2+\widetilde{m}_c^2 \, \delta \left( s-\widetilde{m}_c^2\right)\right]\nonumber\\
&&+\frac{17\langle\bar{q}g_s\sigma Gq\rangle^2}{442368\pi^4}\int dydz \, (y+z)\left[4+\overline{m}_c^2 \, \delta \left( s-\overline{m}_c^2\right)\right] \, ,
\nonumber\\
\widetilde{\rho}^0_{10}(s)&=&\frac{19\langle\bar{q}g_s\sigma Gq\rangle^2}{49152\pi^4}\int_{y_i}^{y_f} dy \, \left[1+\widetilde{m}_c^2 \, \delta \left( s-\widetilde{m}_c^2\right)\right]\nonumber\\
&&+\frac{17\langle\bar{q}g_s\sigma Gq\rangle^2}{221184\pi^4}\int dydz \,\left[3+\overline{m}_c^2 \, \delta \left( s-\overline{m}_c^2\right)\right]\, ,
\end{eqnarray}
where $\int dydz=\int_{y_i}^{y_f}dy \int_{z_i}^{1-y}dz$, $y_{f}=\frac{1+\sqrt{1-4m_c^2/s}}{2}$,
$y_{i}=\frac{1-\sqrt{1-4m_c^2/s}}{2}$, $z_{i}=\frac{y
m_c^2}{y s -m_c^2}$, $\overline{m}_c^2=\frac{(y+z)m_c^2}{yz}$,
$ \widetilde{m}_c^2=\frac{m_c^2}{y(1-y)}$, $\int_{y_i}^{y_f}dy \to \int_{0}^{1}dy$, $\int_{z_i}^{1-y}dz \to \int_{0}^{1-y}dz$ when the $\delta$ functions $\delta\left(s-\overline{m}_c^2\right)$ and $\delta\left(s-\widetilde{m}_c^2\right)$ appear.


\begin{thebibliography}{99}

\bibitem{Gell-Mann-1964} M. Gell-Mann, Phys. Lett. {\bf 8} (1964) 214.

\bibitem{Jaffe-1977}  R. L. Jaffe, Phys. Rev. {\bf D15} (1977) 267.

\bibitem{Strottman-1979} D. Strottman, Phys. Rev. {\bf D20} (1979) 748.

\bibitem{Lipkin-1987} H.  J. Lipkin, Phys. Lett. {\bf B195} (1987) 484.

\bibitem{Maiani-2004}  L. Maiani, F. Piccinini, A. D. Polosa and V. Riquer, Phys. Rev. Lett. {\bf 93} (2004) 212002;
L. Maiani, F. Piccinini, A. D. Polosa and V. Riquer, Phys. Rev. {\bf D71} (2005) 014028.


\bibitem{Jaffe-2003}   R. L. Jaffe and F. Wilczek,  Phys. Rev. Lett. {\bf 91} (2003) 232003.

\bibitem{ReviewAmsler2} C. Amsler and N. A. Tornqvist, Phys. Rept. {\bf 389} (2004) 61.


\bibitem{Weinberg}   S. Weinberg, Phys. Rev. Lett. {\bf 110} (2013) 261601.


\bibitem{Polosa-IJMPLA}  A. Esposito, A. L. Guerrieri, F. Piccinini, A. Pilloni and A. D. Polosa, Int. J. Mod. Phys. {\bf A30} (2014)  1530002.


\bibitem{Brodsky-2014}  S. J. Brodsky, D. S. Hwang and R. F. Lebed, Phys. Rev. Lett. {\bf 113} (2014) 112001.

\bibitem{Maiani-1405}  L. Maiani, F. Piccinini, A. D. Polosa and V. Riquer, Phys. Rev. {\bf D89} (2014) 114010.


\bibitem{Voloshin-2011} A. E. Bondar, A. Garmash, A. I. Milstein, R. Mizuk and M. B. Voloshin, Phys. Rev. {\bf D84} (2011) 054010;
 M. B. Voloshin, Phys. Rev. {\bf D84} (2011) 031502.



\bibitem{Karliner-2015} M. Karliner and J. L. Rosner, Phys. Rev. Lett. {\bf 115} (2015)  122001.


\bibitem{Wang-Tetraquark-DW} Z. G. Wang, Eur. Phys. J. {\bf C74} (2014)  2874;
 Z. G. Wang and T. Huang, Nucl. Phys. {\bf A930} (2014) 63;
  Z. G. Wang, Commun. Theor. Phys. {\bf 63} (2015) 466;
   Z. G. Wang, Commun. Theor. Phys. {\bf 63} (2015) 325.


\bibitem{LHCb-4380} R. Aaij  et al, Phys. Rev. Lett. {\bf 115} (2015) 072001.

\bibitem{mole-penta}  R. Chen, X. Liu, X. Q. Li and S. L. Zhu, Phys. Rev. Lett. {\bf 115} (2015) 132002;
 H. X. Chen, W. Chen, X. Liu, T. G. Steele and S. L. Zhu, Phys. Rev. Lett. {\bf 115} (2015)  172001;
 L. Roca, J. Nieves and E. Oset,  Phys. Rev. {\bf D92} (2015) 094003;
 U. -G. Meissner and J. A. Oller,   Phys. Lett. {\bf B751} (2015) 59;
 J. He, Phys. Lett. {\bf B753} (2016) 547.
 


\bibitem{mole-penta-No}  A. Mironov and A. Morozov, JETP Lett. {\bf 102} (2015)  271.

\bibitem{di-di-anti-penta} L. Maiani, A. D. Polosa and V. Riquer,  Phys. Lett. {\bf B749} (2015) 289;
 V. V. Anisovich, M. A. Matveev, J. Nyiri, A. V. Sarantsev and A. N. Semenova, arXiv:1507.07652;
 G. N. Li, M. He and X. G. He,  JHEP {\bf 1512} (2015) 128.


 \bibitem{di-tri-penta} R. F. Lebed, Phys. Lett. {\bf B749} (2015) 454.

 \bibitem{rescattering-penta} F. K. Guo, U. -G. Meissner, W. Wang and Z. Yang, Phys. Rev. {\bf D92} (2015)  071502;
 X. H. Liu, Q. Wang and Q. Zhao,  arXiv:1507.05359;
 M. Mikhasenko, arXiv:1507.06552.


\bibitem{Test-Penta} Q. Wang, X. H. Liu and Q. Zhao, Phys. Rev. {\bf D92} (2015) 034022;
V. Kubarovsky and M. B. Voloshin, Phys. Rev. {\bf D92} (2015) 031502;
M. Karliner and J. L. Rosner, Phys. Lett. {\bf B752} (2016) 329.


\bibitem{Wang-molecule} Z. G. Wang  and T. Huang, Eur. Phys. J. {\bf C74} (2014)  2891;
  Z. G. Wang, Eur. Phys. J. {\bf C74} (2014)  2963.

 \bibitem{Wang-Octet}  Z. G. Wang, Int. J. Mod. Phys. {\bf  A30} (2015)  1550168.


\bibitem{No-formular} R. D. Matheus, S. Narison, M. Nielsen and J. M. Richard, Phys. Rev. {\bf D75} (2007) 014005;
 F. S. Navarra, M. Nielsen and S. H. Lee, Phys. Lett. {\bf B649} (2007) 166;
   Z. G. Wang, Eur. Phys. J. {\bf C59} (2009) 675;
   Z. G. Wang, Eur. Phys. J. {\bf C63} (2009) 115;
  J. R. Zhang and M. Q. Huang, Commun. Theor. Phys. {\bf 54} (2010) 1075;
  M. Nielsen, F. S. Navarra and S. H. Lee, Phys. Rept. {\bf 497} (2010) 41.


\bibitem{WangHuang-PRD}  Z. G. Wang and T. Huang,  Phys. Rev. {\bf D89} (2014)  054019.


\bibitem{Wang-Tetraquark-DW-2} Z. G. Wang, Mod. Phys. Lett. {\bf A29} (2014) 1450207;
  Z. G. Wang and Y. F. Tian, Int. J. Mod. Phys. {\bf A30} (2015) 1550004.


\bibitem{One-gluon} A. De Rujula, H. Georgi and S. L. Glashow, Phys. Rev.  {\bf D12} (1975) 147;
 T. DeGrand, R. L. Jaffe, K. Johnson and J. E. Kiskis, Phys.  Rev.  {\bf D12} (1975) 2060.


\bibitem{WangDiquark} Z. G. Wang, Eur. Phys. J. {\bf C71} (2011) 1524;
  R. T. Kleiv, T. G. Steele and A. Zhang, Phys. Rev. {\bf D87} (2013) 125018.


\bibitem{WangLDiquark}  Z. G. Wang, Commun. Theor. Phys. {\bf 59} (2013) 451.

\bibitem{WangLambda}  Z. G. Wang, Eur. Phys. J. {\bf C68} (2010) 479.


\bibitem{SVZ79}  M. A. Shifman, A. I. Vainshtein and V. I. Zakharov, Nucl. Phys. {\bf B147} (1979) 385, 448.

\bibitem{PRT85} L. J. Reinders, H. Rubinstein and S. Yazaki, Phys. Rept. {\bf 127} (1985) 1.

\bibitem{HuangShiZhong} Shi-Zhong Huang, "Free particles and fields of high spins" (in chinese), Anhui peoples Publishing House, 2006.

\bibitem{PDG}  K. A. Olive et al, Chin. Phys. {\bf C38} (2014) 090001.

\bibitem{WangHbaryon} Z. G. Wang, Phys. Lett. {\bf B685} (2010) 59;
  Z. G. Wang, Eur. Phys. J. {\bf C68} (2010) 459;
   Z. G. Wang,  Eur. Phys. J. {\bf A45} (2010) 267;
  Z. G. Wang, Eur. Phys. J. {\bf A47} (2011) 81;
    Z. G. Wang, Commun. Theor. Phys. {\bf 58} (2012) 723.


\bibitem{WangPentaquark} Z. G. Wang and T. Huang, arXiv:1508.04189, Z. G. Wang,  arXiv:1509.06436; Z. G. Wang, arXiv:1512.04763.

\bibitem{Nielsen3900} J. M. Dias, F. S. Navarra, M. Nielsen and C. M. Zanetti, Phys. Rev. {\bf D88} (2013) 016004.

\end{thebibliography}
\end{document}